\documentclass[10pt, letterpaper]{article}

\usepackage[margin=1in]{geometry} 
\usepackage[utf8]{inputenc}
\usepackage{amsmath, amsfonts, amssymb, amsthm}
\theoremstyle{definition}
\newtheorem{definition}{Definition}
\newtheorem{assumption}{Assumption}
\usepackage{booktabs}
\usepackage{threeparttable}
\usepackage{multirow}
\usepackage{graphicx}
\usepackage{rotating}
\usepackage{caption}
\usepackage{subcaption}
\usepackage{array}  
\usepackage{url}
\usepackage[authoryear]{natbib}
\usepackage{enumitem}
\usepackage{algorithm}
\usepackage{algorithmic}
\usepackage{hyperref}
\hypersetup{
    colorlinks=true,
    linkcolor=blue,
    citecolor=blue, 
    filecolor=magenta,      
    urlcolor=cyan,
}
\usepackage{authblk} 
\usepackage{url}

\usepackage{cleveref}
\usepackage[title]{appendix}
\usepackage{titling}

\title{Two-Sided Prioritized Ranking: A Coherency-Preserving Design for Marketplace Experiments}

\author{
Mahyar Habibi\thanks{Lyft. \texttt{mhyrhabibi@gmail.com}} \and
Zahra Khanalizadeh\thanks{University of Washington. \texttt{zkhnl@uw.edu}} \and
Negar Ziaeian\thanks{University of Warwick. \texttt{negar.ziaeian-ghasemzadeh@warwick.ac.uk}}
}

\date{March 2026}
\begin{document}

\maketitle

\thispagestyle{empty} 
\begin{abstract}
\noindent Online marketplaces frequently run pricing experiments in environments where users choose from a list of items. In these settings, items compete for users' limited attention and demand, creating interference among items within a list: Changing prices for any item can affect the demand for others, biasing estimates from item-level A/B tests. Besides, a key consideration in pricing experiments is preserving platform coherency across prices and item availability. This requirement rules out experimental designs such as user-level A/B tests as they violate platform coherency. We propose Two-Sided Prioritized Ranking (TSPR) to estimate the total average treatment effect of price changes in such settings. TSPR exploits position bias in ranked search results to create variation in treatment exposure without compromising coherency. TSPR randomizes both users and items and reorders ranked lists, prioritizing treated items for one group of users and untreated items for the other. All users see the same items at consistent prices, but differ in exposure to treatment as they pay disproportionate attention across ranks. In semi-synthetic simulations based on Expedia hotel search data, TSPR outperforms baseline coherency-preserving experiment designs by reducing estimation bias and providing sufficient statistical power.\\


\end{abstract}

\section{Introduction}
\label{sec:intro}
Online platforms such as e-commerce sites and online marketplaces rely heavily on randomized controlled experiments to guide product decisions. These experiments help platforms evaluate changes safely, improve user experience, and increase engagement and sales, while providing timely and credible feedback on new features \citep{kohavi2020trustworthy, bojinov2022online, xia2019safe, xu2018sqr, kohavi2009online}.

Standard experimental designs rely on the Stable Unit Treatment Value Assumption (SUTVA), which rules out spillovers across units \citep{rubin1974estimating, imbens2015causal}. In online marketplaces this assumption is frequently violated. Items compete for users' limited attention and demand within ranked lists, so modifying treated items (for example, via discounts or price increases) can change outcomes for untreated items through substitution or complementarity. Such interference (spillovers, network dependence) has been documented in ridesharing platforms \citep{chamandy2016experimentation} and online pricing experiments \citep{choi2019monetizing}. When interference is ignored, estimates from randomized experiments can be substantially biased \citep{blake2014marketplace, fradkin2019simulation, hudgens2008toward}. For example, if we discount Hotel A in a search result, users who would have booked Hotel B now book A instead, making B appear to perform worse, not because of its true quality, but because of demand spillovers from the treated item.

Under interference, treatment effects depend on how treatment is distributed across units. A natural policy-relevant contrast is the total effect of moving from a fully untreated world to a fully treated world. This contrast is formalized as the global treatment effect in the interference literature \citep{hudgens2008toward, manski2013identification, munro2024treatmenteffectsmarketequilibrium}. 

Interference alone does not preclude credible estimation of global treatment effect, but the experimental designs that would address it are often infeasible in marketplace settings due to operational constraints. User-level randomization, where all items shown to treated users receive treatment, would avoid mixing treated and untreated items within the same query. However, this approach violates what we call \emph{coherency}: the requirement that all users observe the same realized price (or other treatment attribute) for any given item, and that all users retain access to the full catalog of items throughout the experiment.

Price parity is sometimes a legal requirement, and more often a reputational one. Overt price variation is tightly regulated under European competition law \citep{eu_tfeu_art102, ec_application_art102}, which makes user-level price experiments, that show different users different prices, difficult to deploy without raising compliance and reputational concerns. Even when not explicitly illegal, platforms face acute trust and brand risks. Consumer reactions to visible price dispersion are typically strong, and perceived unfairness can dominate any short-run learning benefits \citep{ccakir2025price}. 

Recent reporting describes tests on Instacart in which shoppers were charged different prices for identical items at the same store, a difference that could impact annual spending by as much as $\$1,200$ \citep{kravitz2025instacart_ai_pricing_bibtex}. A September 2025 Consumer Reports survey of 2,240 U.S. adults found that 72 percent of Instacart users opposed such price discrimination for any reason \citep{consumerreports2025_aes_september_omnibus}. Following public scrutiny, Instacart announced in December 2025 that it would end ``item price tests,'' noting that showing different prices for the same item at the same store fell short of customer expectations \citep{instacart2025ending_item_price_tests}. These concerns are longstanding; for instance, Amazon's 2000 DVD pricing experiment triggered immediate backlash and led to necessitating public apologies and refunds \citep{wired2000priceamends}. Disclosing that a price difference is part of an experiment not only risks reputational costs but also undermines internal validity by altering user behavior. Workarounds such as coupon codes or targeted promotions introduce their own confounding incentives and can complicate interpretation of price effects.

A second coherency requirement is \emph{full catalog access}. Many platforms cannot remove items from search results or show different item sets across users without degrading the user experience, distorting substitution patterns, and harming revenue. Designs that vary availability conflate the effect of the intervention with the effect of restricting choice sets, and they may induce strategic seller or user responses that do not reflect business-as-usual behavior.

These two constraints, along with practical considerations, rule out many standard experimental designs. User-level randomization violates coherency by showing different prices to different users. Item-level randomization preserves coherency but suffers from substantial bias under interference \citep{blake2014marketplace}. While cluster randomization can reduce interference by grouping related units \citep{ugander2013graphclusterrandomizationnetwork, ugander2023randomized, holtz2024reducing, saveski2017detecting}, it suffers from low statistical power and requires well-defined clusters that align with spillover patterns, which is often difficult in marketplaces where interactions evolve dynamically, and can be computationally expensive to implement \citep{candogan2023correlated}. 

Recent advances in network interference, such as exposure-based designs, typically require researchers to specify an exposure model or mapping that defines how treatment spillovers propagate across units (e.g., \citet{aronow2017estimating, harshaw2023design}), while analysis-based corrections can debias naive estimates when the platform intermediates spillovers through a known mechanism such as matching \citep{bright2025reducing, chin2019regression}. However, this is difficult to specify and validate in marketplace settings where substitution patterns are context-dependent and high-dimensional. Switchback testing alternates treatment assignments over time for the same units \citep{brown1980crossover, robins1986new, sneider2019experiment, bojinov2023design}. While switchbacks can support causal identification in time-varying environments, frequent treatment fluctuations can confuse users and distort engagement patterns. For salient interventions such as prices, these fluctuations may also create carryover effects that undermine internal validity.

Two-sided randomization (TSR) methods apply randomization on both the user side and the item side \citep{johari2022experimental, bajari2023experimental, li2022interference}. Standard TSR implementations apply treatment only when a treated user interacts with a treated item, which can lead different users to see different versions of the same item (including different prices). This violates the coherency requirement that motivates our setting. This creates a methodological gap: how can platforms credibly estimate treatment effects while maintaining both price parity and full catalog access under interference?

We introduce Two-Sided Prioritized Ranking (TSPR), an experimental design for item-side interventions in ranked-list marketplaces that maintains price parity and full catalog access while addressing interference. Under plausible conditions, we show that TSPR identifies the proportional effect of global treatment.

TSPR exploits a feature of modern marketplaces: centralized recommender systems rank items for each user query, and users exhibit strong position bias, allocating disproportionate attention to top-ranked items \citep{craswell2008experimental, friedberg2022causal, joachims2017accurately, richardson2007predicting}. Empirical evidence shows a steep decline in click probability as an item moves down the ranking. These models imply that observed clicks combine position and relevance effects. The key insight is that while we cannot vary treatment status across users or remove items (coherency), we \emph{can} vary users' exposure to treatment by systematically reordering the ranked list.

Specifically, TSPR randomizes users into two groups and reorders each user's ranked list so that treated items are prioritized at the top for one group and untreated items are prioritized for the other. This induces systematic variation in treatment exposure through position bias while preserving coherency: all users retain access to the same underlying item set, and each item's treatment status remains consistent across users. 

Prior work on interference in ranking experiments often focuses on evaluating or improving ranking algorithms \citep[e.g.,][]{goli2024bias, zhan2024estimating, nandy2021b, ursu2018power}. Our objective is different: we do not treat the recommender system as the object of experimentation, but instead use it as the mechanism through which item-side treatment exposure is shifted while preserving coherency. This differs from approaches that rely on naturally occurring ranking noise as exogenous variation, and from interleaving-style methods that primarily optimize ranking quality rather than deliver coherent item-side interventions.

Using an open-source Expedia hotel search dataset, we estimate behavioral models of click and booking decisions and conduct Monte Carlo simulations to evaluate performance of our method in estimating global treatment effect. TSPR substantially reduces both bias and variance relative to item-level A/B tests, and strongly outperforms cluster-randomized designs on the variance of estimates. These results demonstrate that ranking-based designs can credibly estimate treatment effects in settings where standard methods fail due to interference or operational constraints.

Section~\ref{sec:methodology} formally defines the TSPR design and derives conditions under which it identifies the global treatment effect. Section~\ref{sec:data_simulation} describes the semi-synthetic simulation framework calibrated to Expedia hotel search data. Section~\ref{sec:results} presents Monte Carlo evidence that TSPR substantially reduces bias and improves efficiency relative to Bernoulli-randomized and cluster-randomized baselines. Section~\ref{sec:discussion} interprets these findings, characterizing when TSPR offers the greatest advantages and when simpler designs suffice. Section~\ref{sec:conclusion} concludes.

\section{Methodology} \label{sec:methodology}

\subsection{Two-Sided Prioritized Ranking (TSPR) Experimentation Setup}

We model a two-sided platform as a matching mechanism between a set of queries $q \in Q$, which represent user inputs, and a set of items $i \in I$, which represent the available options. The platform uses a recommender system to compute relevance scores $r_{q,i} \in \mathbb{R}$ for each query–item pair based on attributes of the query and the item, such as user preferences and item features. When a user submits query $q$, the platform ranks all available items in descending order of $r_{q,i}$ and displays the ordered list to the user. After viewing the list, the user may interact with some of the displayed items, and these interactions generate outcomes $y_{q,i}$. For simplicity, we assume that all items begin with outcome value zero and that $y_{q,i}$ takes non-negative real values after user interaction, representing clicks, bookings, or revenue. Because each user submits exactly one query in our setting, we use the terms “user’’ and “query’’ interchangeably.

In this environment, standard item-level A/B testing fails to produce unbiased estimates of treatment effects because items shown together in the same query can affect each other’s outcomes. This violates the Stable Unit Treatment Value Assumption (SUTVA) due to interference between items. Any experimental design for this setting must also satisfy two operational constraints. First, users must retain access to the full catalog of items during the experiment. Second, all users must observe a coherent realization of item treatment status, meaning that every user sees the same version of each item throughout the experiment. These constraints rule out many existing designs and motivate the structure of our Two-Sided Prioritized Ranking approach.

\begin{definition}[\textbf{Coherency}]
A user experience is \textit{coherent} if all users retain access to the same set of items and if every user observes the same treatment status for any given item, independent of their randomized group assignment.
\end{definition}

Due to item-side interference, the effect of a binary treatment $T_i \in \{0,1\}$ on item–query outcomes $y_{q,i}$ depends on how treatments are distributed across items. This motivates our focus on the \emph{global lift} ($\Phi$), which captures both direct effects and spillovers by comparing expected outcomes under full treatment and full control. Because our estimand is defined at the query level, we aggregate item outcomes within each query and work with $Y_q = \sum_i y_{q,i}$. For notational simplicity we omit the query index and write $Y$.

We define global lift as
\begin{equation}
\label{eq:global_lift}
\Phi
\;=\; 
\frac{\mathbb{E}\!\left[\, Y \mid \forall i \in I:\, T_i = 1 \,\right]}
       {\mathbb{E}\!\left[\, Y \mid \forall i \in I:\, T_i = 0 \,\right]} - 1,
\end{equation}
where $I$ is the set of all items. The numerator corresponds to the expected query-level outcome when all items are treated, and the denominator corresponds to the expected outcome when all items are untreated. Since in practice each item can only be in one treatment state at a time, only one of these two quantities is observed, which makes $\Phi$ fundamentally a counterfactual estimand. This estimand is in one-to-one correspondence with the total average treatment effect emphasized in the interference literature \citep{hudgens2008toward, manski2013identification, munro2024treatmenteffectsmarketequilibrium}.

The proposed method rests on several assumptions. First, we assume that items at the top of the listing exert a disproportionate influence on user behavior \citep{craswell2008experimental}, and that this influence declines rapidly with rank. Effective exposure to the treatment therefore depends on the extent to which treated items appear near the top of the ranked list, since these positions receive most of the user’s attention. By strategically altering the ordering of items, we manipulate users’ effective exposure to treated versus untreated items.

Second, the method requires that each query contains a sufficiently large set of relevant items. This ensures that the repositioning scheme can meaningfully increase the exposure of one group of queries to treated items while decreasing it for the other group.

Third, we assume that user-side interference is negligible.
This corresponds to a slack-supply environment in which inventory or availability constraints are not binding over the experiment horizon.
Under slack supply, one user’s actions do not affect item availability for others, and interference arises entirely \emph{within queries}, across items displayed in the same ranked list.
In our model, within-query interference operates through two mechanisms:
(i) limited attention to early ranks and
(ii) unit-demand substitution, since booking one item reduces the probability that other items in the same query are chosen.\footnote{Extending the design to settings with binding capacity constraints or other forms of user-side interference across queries or users is left to future work.}

\begin{figure}[htbp]
    \centering
    \caption{Two-Sided Prioritized Ranking (TSPR) Experimental Design}
    \includegraphics[width=0.85\textwidth]{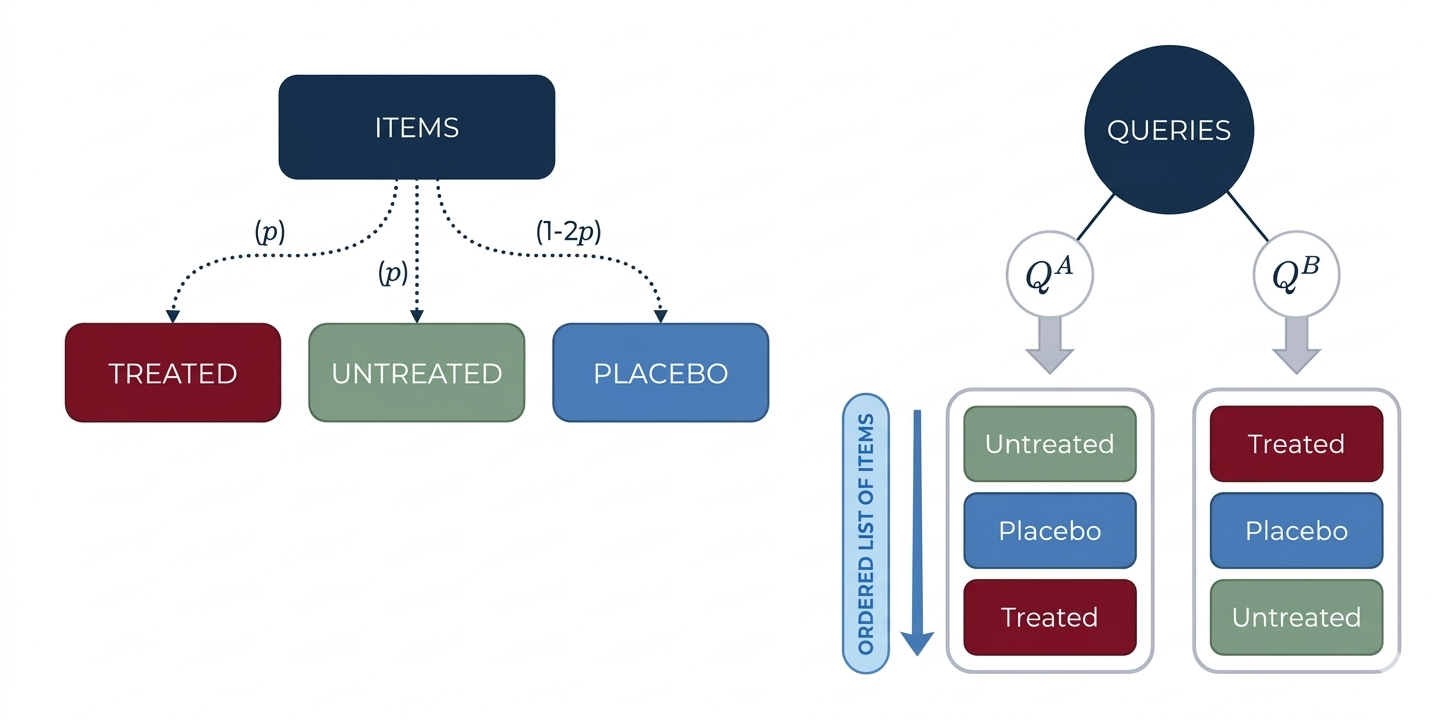}
    \parbox{0.85\textwidth}{\footnotesize{\textit{Notes:} The figure illustrates the TSPR experiment setup. Items are partitioned into three groups, and queries are divided into two subsets. The relevant items for each query are first ordered based on their group-specific priority and then by their relevance score.}}
    \label{fig:setup}
\end{figure}

Our proposed experimental design for estimating total lift is summarized in Table \ref{tab:experiment_setup}, with Figure \ref{fig:setup} illustrating the two-sided randomization scheme and group-specific listing priorities for query results.

\setlength{\aboverulesep}{2pt}
\setlength{\belowrulesep}{2pt}

\begin{table}[htbp]
    \centering
    \caption{Two-Sided Prioritized Ranking (TSPR) Experimental Design}
    \label{tab:experiment_setup}
    \begin{tabular}{>{\raggedright\arraybackslash}p{0.85\textwidth}}  
        \toprule
        \textbf{TSPR Experiment Setup}\\
        \midrule
    \begin{enumerate}[leftmargin=*, topsep=2pt, itemsep=2pt]            \item Set the probability of receiving treatment for an item $p < 0.5$.

            \item Randomize items into Treated, Untreated, and Placebo subsets with probabilities $p$, $p$, and $1-2p$, respectively. Apply the treatment only to the Treated group.

            \item For each incoming query $q$:
            \begin{enumerate}[label=3.\arabic*., leftmargin=*, topsep=2pt, itemsep=2pt]
                \item Randomly assign $q$ to $Q^A$ or $Q^B$  and set the item priorities as follows:
                \begin{itemize}
                    \item If $q \in Q^A$: 1-Untreated, 2-Placebo, and 3-Treated.
                    \item If $q \in Q^B$: 1-Treated, 2-Placebo, and 3-Untreated.
                \end{itemize}
                
                \item Rank items primarily by priority (ascending) and secondarily by relevance score (descending).
            \end{enumerate}
        \end{enumerate}\\
        \bottomrule
    \end{tabular}
\end{table}

As outlined in Table \ref{tab:experiment_setup}, after specifying the experiment intensity parameter $p$, we begin by partitioning items into three subsets: Treated, Untreated, and Placebo, with probabilities $p$, $p$, and $1 - 2p$, respectively. Only items in the Treated subset receive the intervention. The inclusion of a Placebo subset is essential for maintaining balance in the experiment.

In marketplace experiments, the probability of assignment to either treatment or control is typically well below 0.5, often on the order of a few percent, in order to limit opportunity costs and to mitigate potential negative effects on user experience if the new feature performs worse than the existing one \citep{ha2020counterfactual}. Without a Placebo subset, the Untreated subset would be substantially larger than the Treated subset. This would create an asymmetric effect in step 3 of our design. In particular, for queries in $Q^A$, where non-treated items are prioritized, the larger Untreated pool would produce top-ranked items of higher average quality than the top-ranked items drawn from the smaller Treated pool shown to $Q^B$. Such an imbalance would cause the recommender system modification to affect the two query groups differently, confounding the estimation of the intervention’s effect.

The Placebo subset prevents this imbalance by ensuring that the Treated and Untreated subsets are of comparable size. As a result, the expected match quality of top-ranked items is similar across the two user groups, which allows the variation induced by the prioritization scheme to isolate the treatment effect rather than reflect differences in pool size or quality.

Placebo items also create a buffer between treated and untreated items in the ranked list. This separation sharpens the interpretation of rank depth as treatment exposure in our partial-outcome contrasts: at small depths, outcomes are driven primarily by exposure to the prioritized block rather than by immediate mixing of treated and untreated items. As a result, placebo reduces contamination of the control arm from treated items and improves the signal-to-noise ratio of the within-depth contrasts that identify lift.

In the next step, incoming queries are randomized into $Q^A$ or $Q^B$ with equal probability. Item priorities are then assigned so that queries in $Q^A$ receive items in the order Untreated, Placebo, Treated, while queries in $Q^B$ receive items in the reverse order: Treated, Placebo, Untreated. This prioritization induces systematic differences in exposure to treated items across the two query groups.

\subsection{Theoretical Setup and Estimation Framework}

This section introduces the analytical framework used throughout the paper. We begin by outlining the setup and notation, then define the estimand of interest that captures treatment effects under varying ranking and attention conditions. We next formalize the identifying assumptions required for consistent estimation and describe the estimator that operationalizes these ideas in practice.

The parameter $\Phi$ captures the relative (multiplicative) effect of treatment. It is defined as the proportional lift in the total outcome under universal treatment (all items treated) relative to universal control (all items untreated) (Equation~\ref{eq:global_lift}). Thus, $\Phi$ represents the percentage change in expected total outcomes when all items are treated.

A TSPR experiment is characterized by the set of treatment-prioritized queries $Q^B$, the set of control-prioritized queries $Q^A$, the set of items $\mathcal{I}$, the treatment intensity $p$, the treatment type $T$, and the randomization and re-ranking scheme implemented according to Algorithm~\ref{tab:experiment_setup}. Let $y_{q,i}$ denote the user/query $q$'s outcome of the item displayed at rank $i$.

\begin{definition}[\textbf{Partial Outcome}]
The \emph{partial outcome}, denoted $Y^{l}_q = \sum_{i=1}^{l} y_{q,i}$, is the cumulative outcome for query $q$ over the first $l$ listed items. As item-level outcomes are assumed to be non-negative ($y_{q,i} \geq 0$), $\mathbb{E}[Y^{l}_q]$ is non-decreasing in $l$.
\end{definition}

For notational simplicity, we drop the query index $q$ and refer to query-level outcomes as $Y$. All expectations are taken over queries within a given experimental arm.

\begin{definition}[\textbf{Attention Function}]
\label{def:attention}
The attention function, $F(l)$, is defined so that under a given ranking: $\mathbb{E}[Y^{l}] = F(l)\mathbb{E}[Y]$, where $F : \mathbb{N}\to(0,1]$ is increasing and concave, and $F(l) \rightarrow 1$ as $l\rightarrow\infty$.
\end{definition}

\begin{assumption}[\textbf{Attention and Treatment Separability}]
\label{ass:separability}
If all items are treated, the treatment affects the level but not the shape of the attention function.
\end{assumption}

Assumption \ref{ass:separability} implies that if treatment were rolled out to all items but still under original recommender system, the expected partial outcome would satisfy
\begin{equation}
\label{eq:partial_lift}
\mathbb{E}[Y^{l} \mid full \; treatment]
= (1 + \Phi) \cdot F(l) \cdot \mathbb{E}[Y \mid no \; treatment].
\end{equation}

We now characterize how moving from the platform’s original ranking to the TSPR ranking experiment (Table~\ref{tab:experiment_setup}) changes expected partial outcomes. There are two channels. First, re-ranking can change user attention, meaning how attention is allocated across positions (for example, time spent evaluating items and clicks). Second, it can change outcomes through the treatment itself. Assumptions~\ref{ass:multiplicative_distortion} and \ref{ass:symmetric_distortion} formalize the distortion induced by re-ranking and how it affects partial outcomes under TSPR. Assumptions~\ref{ass:partial_treatment_effect} and \ref{ass:nuisance_treatment_effect} then describe how the partial treatment exposure queries receive affects partial outcomes, both when treated items are prioritized at the top for group $B$ queries and when treated items are down-ranked for group $A$ queries.

In a TSPR experiment, the platform perturbs its baseline relevance ordering to induce exogenous variation in exposure. Such changes alter how user attention is distributed across the list. Because the baseline recommender maximizes outcomes by favoring highly relevant items near the top, these perturbations generally lower total and partial outcomes. When items that would appear lower under the platform’s baseline ranking are moved into early positions, users may click less, search less deeply, or abandon sooner. The magnitude of this perturbation is governed by the treatment assignment probability $p$ from the experimental design: larger $p$ implies a larger expected deviation from the platform’s baseline ordering within early ranks. We capture this channel by allowing the baseline attention function $F(l)$ to be attenuated under TSPR, and denote the distorted attention function by $D(l;p)$.

\begin{assumption}[\textbf{Multiplicative Distortion}]
\label{ass:multiplicative_distortion}
TSPR re-ranking distortion attenuates attention multiplicatively:
\[
D(l;p) = d(l;p)\,F(l),
\]
where $d(l;p)\in(0,1]$ is a depth-$l$ attenuation factor that depends on the treatment assignment probability $p$.
\end{assumption}

\begin{assumption}[\textbf{Symmetric Distortion}]
\label{ass:symmetric_distortion}
Conditional on the treatment assignment probability $p$, the re-ranking attenuation is identical across experimental arms. That is, for all depths $l$,
\[
d_A(l;p)=d_B(l;p)\equiv d(l;p).
\]
Equivalently, TSPR induces the same expected attention distortion in arms $A$ and $B$.
\end{assumption}

Assumption~\ref{ass:symmetric_distortion} is motivated by the symmetry of the TSPR design. Items are randomly assigned to Treated, Untreated, and Placebo labels, independently of their baseline relevance. As a result, the distributions of baseline relevance among Treated and Untreated items are identical in expectation. The two query arms then apply mirror-image block prioritization rules: arm $B$ promotes the Treated block while arm $A$ promotes the Untreated block, and in both arms items are otherwise the same and only re-ordered within a fixed candidate set. When within-block ordering follows the platform’s baseline ranking, the primary source of perturbation is the block swap itself, whose magnitude is governed by the treatment assignment probability $p$. Under these conditions, the expected quality of the top-$l$ positions, and therefore the induced attenuation in attention, is the same across arms, implying $d_A(l;p)=d_B(l;p)$ for all $l$.

Equation \eqref{eq:partial_lift} characterizes partial outcomes under full treatment. Under TSPR, however, treatment is applied to only a small subset of items, but the re-ranking scheme uses position bias to maximize exposure to treated items for queries in $Q^B$ and minimize exposure for queries in $Q^A$. Partial treatment and ranking distortion therefore require a more general formulation.

Invoking Assumptions~\ref{ass:separability}, \ref{ass:multiplicative_distortion}, and \ref{ass:symmetric_distortion}, and writing $d(l)$ for the distortion function at a fixed treatment probability $p$, the expected partial outcome at rank $l$ for a query assigned to group $B$ under TSPR satisfies
\begin{equation}
\label{eq:full_t_TSPR}
    \mathbb{E}[Y^{l} \mid \text{Full treatment}, \text{TSPR}]
    =
    (1+\Phi)\, d(l)\, F(l)\,
    \mathbb{E}[Y \mid \text{No treatment}, \text{original ranking}].
\end{equation}
Similarly, for a query in group $A$ in TSPR,
\begin{equation}
\label{eq:no_t_TSPR}
    \mathbb{E}[Y^{l} \mid \text{No treatment}, \text{TSPR}]
    =
    d(l)\, F(l)\,
    \mathbb{E}[Y \mid \text{No treatment}, \text{original ranking}].
\end{equation}

We now introduce two functions, $\tau(\cdot)$ and $\nu(\cdot,\cdot)$, that characterize how treatment exposure interacts with ranking in treatment-dominated ($Q^B$) and control-dominated ($Q^A$) listings. The function $\tau(\cdot)$ captures the \emph{scaling of treatment effects} when treated items fill the top positions in group $B$, reflecting substitution or complementarity across these items. The function $\nu(\cdot,\cdot)$ captures the \emph{contamination effect} for group $A$, where a small number of treated items may appear in lower ranks and influence expected outcomes.

Building on equation~\ref{eq:full_t_TSPR}, for a query in group $B$ with $l\leq n_b$ treated items at the top:
\begin{equation}
\label{eq:partial_lift_B}
\mathbb{E}[Y^{l}_B \mid \text{TSPR}]
= (1 + \tau(n_b)\Phi)\; d(l) \; F(l) \; \mathbb{E}[Y\mid \text{No treatment}].
\end{equation}

\begin{assumption}[\textbf{Partial Treatment Effect}]
\label{ass:partial_treatment_effect}
$\tau:\mathbb{N}\to\mathbb{R}_+$ satisfies $\tau(l)\to 1$ as $l\to\infty$. 
The function may converge from below ($\tau(1)<1$) with a concave shape in $l$, 
from above ($\tau(1)>1$) with a convex shape in $l$, 
or be constant ($\tau(\cdot)=1$).
\end{assumption}

Assumption~\ref{ass:partial_treatment_effect} ensures that partial lift has the same sign as the full-treatment effect, and that $\phi(l)=\Phi\,\tau(l)$ converges to $\Phi$ as the treated block grows.
The sign of $\tau(1)-1$ is a reduced-form summary of net interference at shallow depths.
When items are substitutes, treating a small block at the top \emph{amplifies} the per-item effect: with $\Phi<0$, for instance, the single treated item at rank~1 loses demand to the many untreated items below it, making $\phi(1)$ more negative than~$\Phi$ and hence $\tau(1)>1$.
Conversely, if items are complements, where the treatment effect grows with the number of co-treated items, treating a single item in isolation produces a smaller effect than treating the full catalog, yielding $\tau(1)<1$.
As the treated block expands and fewer untreated items remain to absorb substitution (or contribute complementarities), $\tau(l)$ declines toward~1.

In our main application (an Expedia-like marketplace), items are substitutes in expectation: booking one hotel forgoes others in the same query.
This substitution channel implies $\tau(1)>1$, with $\tau(l)$ decreasing toward~$1$ as $l$ grows.

For a query in group $A$, with $n_u$ untreated items, $n_p$ placebo items, and $n_a$ treated items appearing at the tail of the list, we model partial outcomes for $l\leq n_u+n_p$ as:
\begin{equation}
\label{eq:partial_lift_A}
\mathbb{E}[Y^{l}_A \mid \text{TSPR}]
= (1 + \nu(n_u + n_p, n_a)\Phi) \, d(l) \, F(l) \, \mathbb{E}[Y\mid \text{No treatment}].
\end{equation}
Here $\nu(\cdot) \in [0,1)$ represents the fractional effective exposure to treatment reaching the top of the control-prioritized arm, scaling outcomes in the direction of $\Phi$.

\begin{assumption}[\textbf{Contamination Effect}]
\label{ass:nuisance_treatment_effect}
The nuisance function
$\nu : \mathbb{Z}\times\mathbb{Z}\rightarrow [0,1)$
is decreasing in the number of untreated and placebo items and increasing in the number of treated items. It satisfies $\nu(\cdot,0)=0$, and $\nu\to 0$ as exposure to treated items becomes negligible.
\end{assumption}

We now define the partial lift of treatment at block size $l$ by conditioning on queries whose prioritized block contains exactly $l$ items, that is, $n_b=l$ for queries in $Q^B$ and $n_u=l$ for queries in $Q^A$:
\begin{equation}
\label{eq:partial_prop_change_general}
1 + \phi(l)
= \frac{\mathbb{E}[Y^l_B]}{\mathbb{E}[Y^l_A]}
= \frac{1 + \tau(n_b)\Phi}{1 + \nu(n_u+n_p,n_a)\Phi}.
\end{equation}

In practice, the contamination term $\nu(\cdot)$ is small whenever treated items in the tail of group~$A$ listings receive little effective attention, either because attention decays sharply with rank, because the list is long, or because $p$ is small so that the placebo buffer separating untreated and treated blocks is wide.

We therefore adopt the approximation $\nu \approx 0$ for the
remainder of the analysis, reducing the partial lift to the
single-function form $\phi(l) = \Phi\,\tau(l)$. This simplification is what allows us to recover $\Phi$ from
the depth profile of $\widehat{\phi}(l)$ using either the parametric or nonparametric estimators described below.

\subsection{Estimation}
\label{sec:estimation}

Given data from a TSPR experiment, we first construct the empirical partial lift at each depth~$l$.
For each block size~$l$, we compare queries from $Q^B$ that have exactly $l$ Treated items in the top positions
to queries from $Q^A$ that have exactly $l$ Untreated items in the top positions,
computing the partial outcome up to position~$l$ in both cases.
The empirical partial lift is
\begin{equation}
\label{eq:phi_hat}
\widehat{\phi}(l)
\;=\;
\frac{\widehat{\mathbb{E}}[Y^l_B]}{\widehat{\mathbb{E}}[Y^l_A]}
\;-\; 1,
\end{equation}
where $\widehat{\mathbb{E}}[Y^l_B]$ and $\widehat{\mathbb{E}}[Y^l_A]$
denote the sample means of the partial outcome in the treatment-prioritized and
control-prioritized arms, respectively, conditional on block size~$l$.

With empirical values $\widehat{\phi}(l)$ for a range of depths $l = 1,\dots,L$, and under standard regularity conditions (e.g., sufficient support across~$l$), we recover the global lift~$\Phi$ using either a parametric or a nonparametric approach. Both methods exploit the relationship $\phi(l) = \Phi\,\tau(l)$, where $\tau(l) \to 1$ as $l \to \infty$, so that $\phi(l)$ converges to the full-treatment effect as depth increases.

\subsubsection{Parametric Estimation: Weighted Least Squares}
\label{sec:estimation_parametric}

We impose a parsimonious parametric form on the depth function
that satisfies the qualitative restrictions derived above; namely, that $\tau(l)$ is smooth, positive, and converges to unity as~$l$ grows:
\begin{equation}
\label{eq:tau_simple}
\tau(l) \;=\; 1 + \frac{1}{l}.
\end{equation}
This specification\footnote{While we also explored a generalized specification $\tau(l) = 1 + \alpha/l$ where $\alpha$ is estimated as a free parameter, we found that this added flexibility yielded no significant improvement in bias reduction; consequently, we maintain the simpler, parameter-free form for all subsequent results. In general, more flexible functional forms can be used depending on the empirical setting.} has no free parameters beyond the global lift $\Phi$ itself. At $l=1$, $\tau(1) = 2$: when only a single treated item occupies the top position, the partial lift is twice the full-treatment effect, reflecting the concentrated exposure of the treatment block. As $l$ increases, $\tau(l)$ declines monotonically toward $1$, so that $\phi(l) \to \Phi$. 

Since the model $\phi(l) = \Phi\,\tau(l)$ is linear in~$\Phi$ for known~$\tau$,
the estimator admits a closed-form weighted least squares solution.
We minimize
\begin{equation}
\label{eq:wls_objective}
Q(\Phi)
\;=\;
\sum_{l=1}^{L}
w(l)
\left[
\widehat{\phi}(l) \;-\; \Phi\,\tau(l)
\right]^2,
\end{equation}
where $w(l)$ is a nonnegative precision weight proportional to the
number of queries contributing to the $l$-th partial outcome.
Setting the first-order condition to zero yields
\begin{equation}
\label{eq:wls_phi_hat}
\widehat{\Phi}_{\mathrm{wls}}
\;=\;
\frac{\displaystyle\sum_{l=1}^{L} w(l)\;\tau(l)\;\widehat{\phi}(l)}
     {\displaystyle\sum_{l=1}^{L} w(l)\;\tau(l)^2}.
\end{equation}
This is equivalent to dividing each $\widehat{\phi}(l)$ by its
model-implied scaling factor~$\tau(l)$ and taking a precision-weighted average of the resulting depth-specific estimates of~$\Phi$. Standard errors are computed via bootstrap resampling at the query level.

\subsubsection{Nonparametric Estimation: Isotonic Regression}
\label{sec:estimation_nonparametric}

As a robustness check that avoids imposing a specific functional form on $\tau(l)$, we estimate $\Phi$ using isotonic regression \citep{barlow1972statistical}. This approach follows recent work in marketplace analytics that utilizes shape restrictions to correct for rank-based biases \citep[e.g.,][]{goli2024bias}. While \citeauthor{goli2024bias} employ isotonic regression to improve recommender system accuracy, we leverage it here to nonparametrically recover the global lift $\Phi$ from item-level interventions. The key shape restriction is that $|\phi(l)|$ is monotonically decreasing in~$l$:
as more items in the top block receive treatment (or control),
the partial lift converges toward the full-treatment effect.
When $\Phi < 0$, this implies that $\phi(l)$ is increasing;
when $\Phi > 0$, $\phi(l)$ is decreasing.

We fit a weighted isotonic regression of $\widehat{\phi}(l)$ on~$l$, using the same precision weights~$w(l)$ as above,
subject to the appropriate monotonicity constraint.
Block sizes with fewer than a minimum number of queries in either arm are excluded prior to fitting to reduce noise from imprecisely estimated cells.

Let $\widetilde{\phi}(l)$ denote the isotonic-regression fitted values. Because $\phi(l) \to \Phi$ as~$l$ grows, we estimate the global lift as a weighted average of the fitted values at large depths:
\begin{equation}
\label{eq:iso_phi_hat}
\widehat{\Phi}_{\mathrm{iso}}
\;=\;
\frac{\displaystyle\sum_{l \in \mathcal{L}_{\mathrm{top}}} w(l)\;\widetilde{\phi}(l)}
     {\displaystyle\sum_{l \in \mathcal{L}_{\mathrm{top}}} w(l)},
\end{equation}
where $\mathcal{L}_{\mathrm{top}}$ contains the largest observed block sizes, selected so that the fitted values have converged close to $\Phi$. The threshold defining $\mathcal{L}_{\text{top}}$ balances two competing forces: restricting to the largest block sizes ensures that the fitted values $\widetilde{\phi}(l)$ have converged close to $\Phi$, while including more block sizes improves precision.

\section{Data and Simulation Setup} \label{sec:data_simulation}

To illustrate our methodology, we use an open-source dataset of hotel search impressions from Expedia \citep{expedia-personalized-sort}. The data capture consumer queries and their subsequent search behavior over an eight-month period. Our training and calibration sample consists of a 20\% subset of the cleaned data, comprising nearly 2 million observation-level records across approximately 80,000 unique search impressions.

Consumers interact with the platform in three stages. First, consumers initiate queries by specifying trip details. Second, they receive a ranked list of hotel results. A key feature of this dataset is the experimental variation in ranking: approximately 30\% of search impressions were randomly sorted, while the remainder followed the platform's original relevance-based recommender system. This variation allows us to disentangle the causal effect of display position from item relevance. Finally, users engage by clicking on hotels to view details and may subsequently complete a booking.

To evaluate our experimental design, we implement Monte Carlo simulations that replicate this two-sided marketplace. We model user interactions as a function of a latent utility $v_{ij}$. The platform's relevance score $r$ is modeled as $r = v + \epsilon$, where $\epsilon \sim N(0, \sigma^2)$. While the original ranking is sorted decreasing in $r$, the random ranking allows for unbiased estimation of position effects. 

Table \ref{tab:data_summary_stats} presents summary statistics for the search impressions used in our analysis, highlighting the baseline differences in performance between the ranking mechanisms.

\begin{table}[ht]
\caption{Summary Statistics of Search Impressions}
\centering
\begin{tabular}{lrrrrrr}
\hline
                        & Mean  & Median    & Min & Max \\
\hline
Randomized Ranking (Yes=1)   & 0.30  & 0   & 0   & 1   \\
Total Hotels per Impression  & 24.56 & 29  & 4   & 33  \\
Clicks per Impression        & 1.11  & 1   & 1   & 30  \\
Bookings per Impression       & 0.69  & 1   & 0   & 1   \\
\quad --- Random ranking      & 0.13  & 1   & 0   & 1   \\
\quad --- Original ranking      & 0.93  & 1   & 0   & 1   \\
\hline
\end{tabular}
\label{tab:data_summary_stats}
\end{table}

\subsection{Click Model}
\label{subsection:click_model}

Click behavior is modeled using a logistic function incorporating rank-based attention and sequential behavior. For each item $j$ at position $p$ for user $i$, the probability of a click is:
\begin{equation}
\label{eq:click_model}
P(\text{click}_{ij}) = \text{logit}^{-1}\Big(
    \beta_1 p_{ij} + \beta_2 p_{ij}^{2} + \beta_3 \text{prevclicks}_{i} + \beta_4 \mathbf{1}[\text{prevclicks}_{i} > 0] + \beta_5 v_{ij} + \beta_0
\Big)
\end{equation}

We estimate these parameters in two stages to resolve the endogeneity of position in relevance-sorted results. In the \textit{first stage}, we use only the randomly sorted subset to estimate the position coefficients ($\beta_1, \beta_2$) and click-history effects ($\beta_3, \beta_4$). Because positions are assigned randomly, these estimates represent pure attention effects. In the \textit{second stage}, we fix these coefficients as an offset and estimate the utility coefficient $\beta_5$ and intercept $\beta_0$ on the full mixed sample. The resulting parameters, shown in Table \ref{tab:click_params}, show a clear initial decline in attention (negative $\beta_1$) and a strong negative pressure on subsequent clicks once an initial click has occurred ($\beta_4$).

\begin{table}[htbp]
\centering
\caption{Estimated Click Model Parameters}
\begin{tabular}{llr}
\hline
Parameter & Variable & Estimate \\
\hline
$\beta_1$ & Position ($p$) & -0.0697 \\
$\beta_2$ & Position Squared ($p^2$) & 0.0025 \\
$\beta_3$ & Previous Clicks Count & 0.6550 \\
$\beta_4$ & Has Clicked (Indicator) & -3.6033 \\
$\beta_5$ & Latent Utility ($v$) & 0.0897 \\
$\beta_0$ & Intercept & -1.7873 \\
\hline
\end{tabular}
\label{tab:click_params}
\end{table}

\begin{figure}[htbp]
  \centering
  \caption{Click and Booking Models}
  \includegraphics[width=\textwidth]{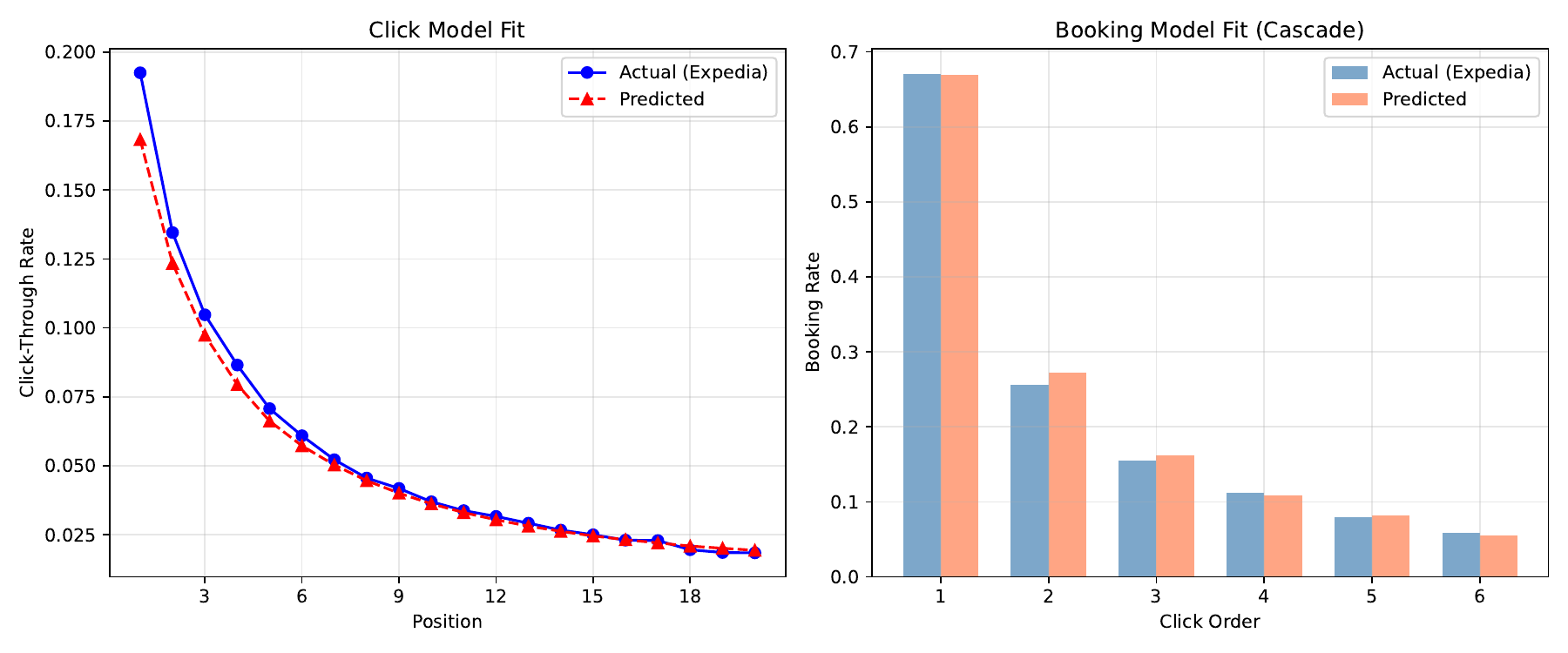}
  \parbox{0.85\textwidth}{\footnotesize{\textit{Notes:} \emph{Left:} Actual vs.\ predicted click-through rate by display position.  \emph{Right:} Actual vs.\ predicted booking rate by click order under the cascade booking model.  Both panels are computed on a hold-out sample not used in estimation.}}
  \label{fig:click_booking_fit}
\end{figure}

The resulting estimates (Table \ref{tab:click_params}) reveal a nuanced search process. The large negative coefficient for the click indicator ($\beta_4 = -3.60$) captures the expected ``satisficing'' effect, where any initial click significantly lowers the marginal probability of further search. However, the positive coefficient for the cumulative click count ($\beta_3 = 0.65$) identifies latent searcher heterogeneity: conditional on not stopping, users with higher click counts exhibit a higher baseline propensity for exhaustive search. This specification ensures the simulation reflects a marketplace populated by both ``quick-search'' and ``high-intensity'' consumers. Figure~\ref{fig:click_booking_fit}~(left) shows that our click model closely fits the click-through-rate based on item position to the real Expedia click impressions data.

\subsection{Booking Model}
\label{subsection:booking_model}

Conditional on the set of clicked items $C_i$, user $i$ either books one item or takes the outside option (no booking).  We model this choice with a multinomial logit that includes a click-order effect to capture cascade behavior in purchasing decisions.  The probability of booking item $k \in C_i$ is
\begin{equation}
\label{eq:booking_model}
P(\text{book}_{ik} \mid C_i) \;=\;
  \frac{\exp\!\big(\gamma_1 \, v_{ik} + \gamma_2 \, \text{click\_order}_{ik} + \gamma_0\big)}
       {1 + \displaystyle\sum_{k' \in C_i} \exp\!\big(\gamma_1 \, v_{ik'} + \gamma_2 \, \text{click\_order}_{ik'} + \gamma_0\big)},
\end{equation}
where $\text{click\_order}_{ik}$ records the sequential position in which item $k$ was clicked (1\,=\,first clicked, 2\,=\,second, etc.) and the ``$1$'' in the denominator represents the outside option.  The coefficient $\gamma_2$ captures the empirical regularity that earlier-clicked items are more likely to be booked, consistent with directed search in which users click the most promising options first.  All parameters are estimated by maximum likelihood on the subsample of clicked items.

\begin{table}[htbp]
\centering
\caption{Estimated Cascade Booking Parameters}
\begin{tabular}{llr}
\hline
Parameter & Variable & Estimate \\
\hline
$\gamma_1$ & Latent Utility ($v$) & 0.4396 \\
$\gamma_2$ & Click Order & -0.1774 \\
$\gamma_0$ & Intercept & 0.4271 \\
\hline
\end{tabular}
\label{tab:booking_params}
\end{table}

The negative coefficient for click order ($\gamma_2$) in Table \ref{tab:booking_params} confirms a ``first-mover'' advantage in search, where items discovered earlier in the process are more likely to be converted. A likelihood ratio test strongly rejects the standard multinomial logit (without click order) in favor of the cascade specification ($p < 0.001$; see Appendix~\ref{appendix:booking_model_comparison}), so we adopt the cascade model for all main results.  Figure~\ref{fig:click_booking_fit}~(right) confirms that the cascade model closely matches observed booking rates by click order.

\subsection{Treatment and Interference}
\label{subsection:treatment}

Treatment is introduced as a constant shift $\delta$ in the latent utility: $v_{ij}^* = v_{ij} + \delta T_{ij}$. This shift propagates through both the click and booking stages. Crucially, as shown in Equation \ref{eq:booking_model}, an increase in the utility of a treated item increases its own booking probability while simultaneously decreasing the probability for all other items in $C_i$. This within-query substitution is the primary source of interference we aim to address with the TSPR design.

\section{Baselines and Results} \label{sec:results}

We conduct counterfactual simulations for 20{,}000 queries using the estimated models of click and booking behavior. To establish a simulated ground truth for lift, we simulate the marketplace under two extreme scenarios: one in which no items receive treatment and one in which all items are treated. The treatment enters as a constant reduction in the latent utility of an item, which represents the effect of a platform-wide price or markup increase and implies a 6.9 percent decline in bookings under full treatment. The recommender system is held fixed in both simulations. The resulting proportional change in total bookings serves as the benchmark against which we evaluate the lift estimates produced by each experimental design.

We then implement our Two-Sided Prioritized Ranking (TSPR) experimental design, described in Table~\ref{tab:experiment_setup} and estimate the total lift using both the parametric (Section~\ref{sec:estimation_parametric}) and non-parametric (Section~\ref{sec:estimation_nonparametric}) approaches. To compare performance of our experimental design and two estimators, we use bernoulli-randomized and cluster-randomized experiments, both common benchmarks for marketplace experiments. All baselines target the same estimand $\Phi$ defined in Equation~\ref{eq:global_lift}; differences in performance reflect interference-induced bias and variance, not differences in the target parameter.
The main results are under treatment probability $p=0.25$ but we show sensitivity to choice of $p$ in Appendix~\ref{app:sensitivity_p}.

\subsection{Performance Baseline: Bernoulli-Randomized A/B Testing}

\begin{figure}[htbp]
    \centering
    \caption{Item-side A/B v. TSPR Design}
    \vspace{0.5em}
    
    \begin{minipage}{0.85\textwidth}
        \centering
        
        \begin{minipage}[t]{0.31\linewidth}
            \centering
            \subcaption*{\textbf{(a) Item-side A/B test}}
            \vspace{0.5em}
            \includegraphics[width=\linewidth]{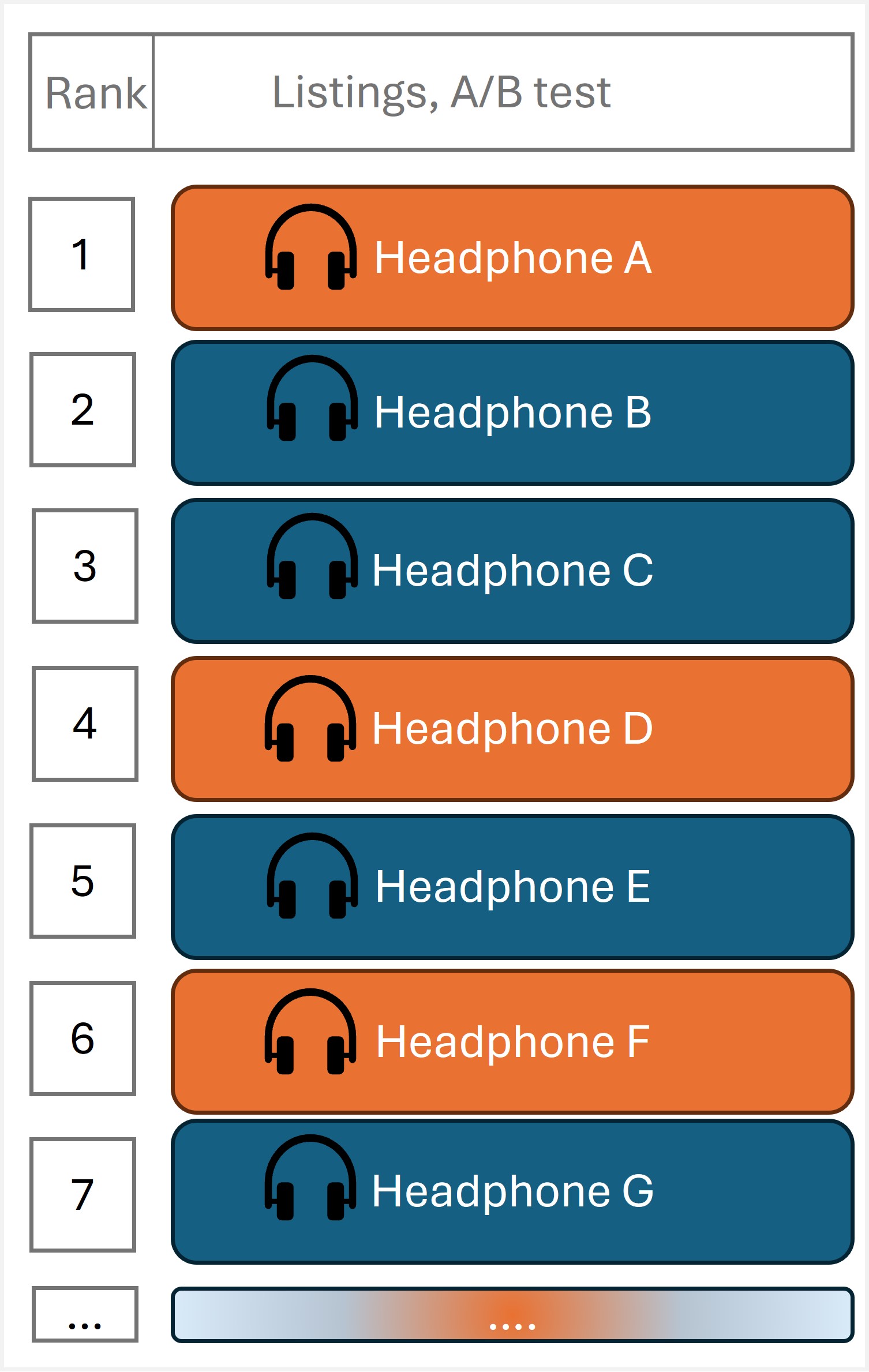}
        \end{minipage}
        \hfill
        \begin{minipage}[t]{0.31\linewidth}
            \centering
            \subcaption*{\textbf{(b) TSPR: \\ \; Treatment arm ($Q^B$)}}
            \vspace{0.5em}
            \includegraphics[width=\linewidth]{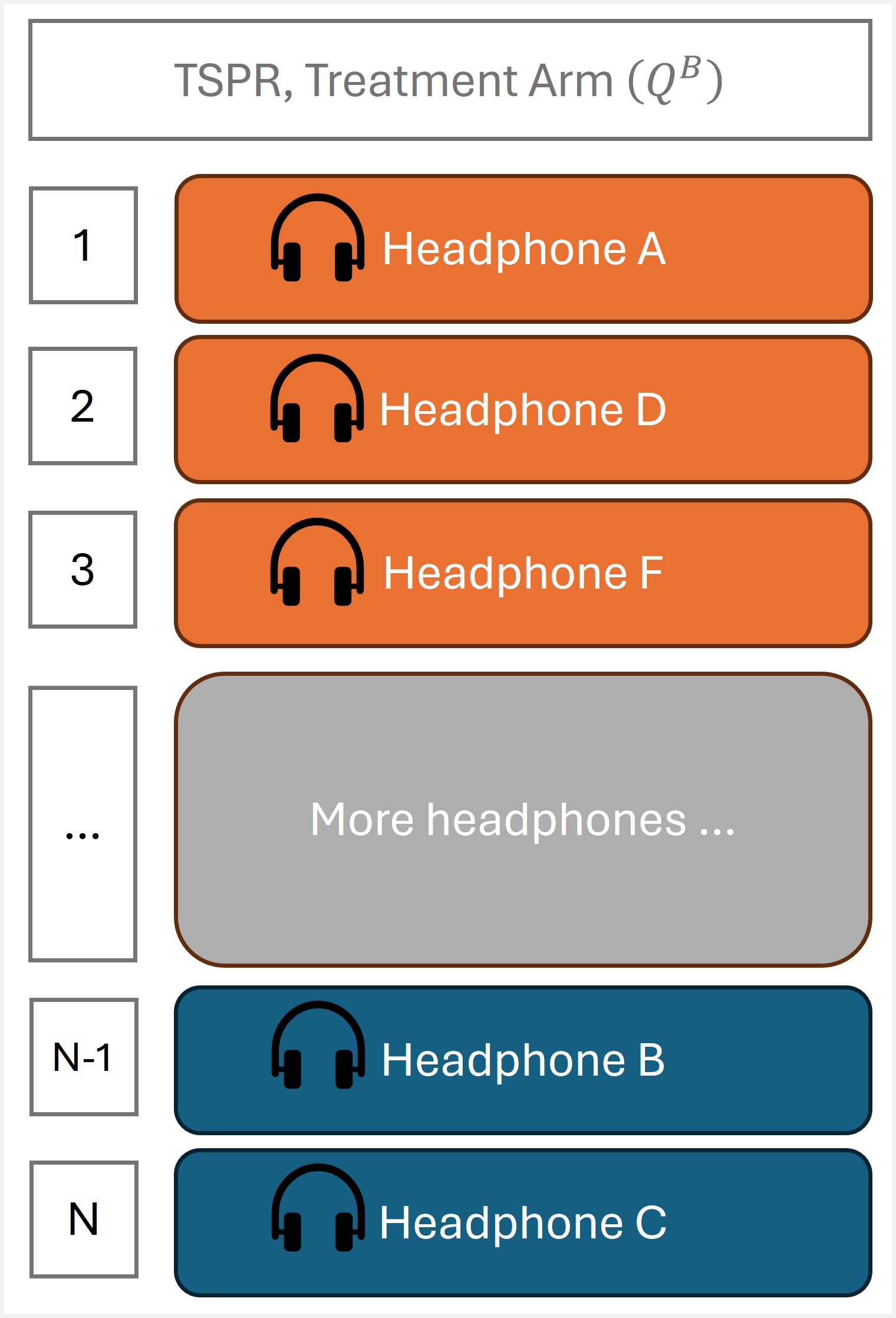}
        \end{minipage}
        \hfill
        \begin{minipage}[t]{0.31\linewidth}
            \centering
            \subcaption*{\textbf{(c) TSPR: \\ \; Control arm ($Q^A$)}}
            \vspace{0.5em}
            \includegraphics[width=\linewidth]{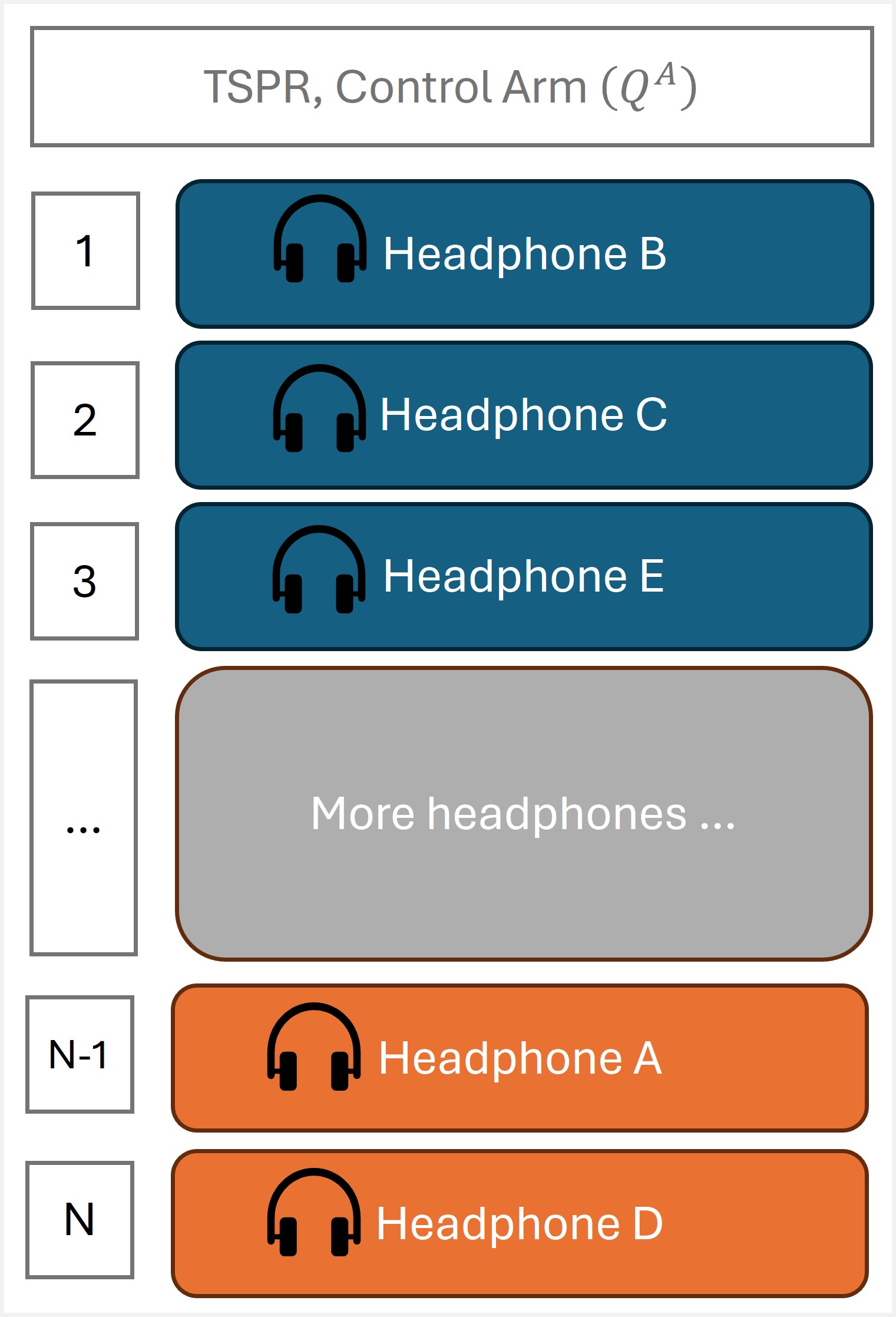}
        \end{minipage}
        
        \vspace{1em} 
        
        \parbox{\linewidth}{\footnotesize{\textit{Notes:}   The figure provides an illustrative comparison of an item-side A/B test and the Two-Sided Prioritized Ranking (TSPR) design. Orange items indicate treated items, blue items indicate untreated items, and gray items indicate placebo items.
        Panel (a) shows a Bernoulli-randomized item-side experiment in which treated and untreated items are interleaved throughout the ranking, generating within-list interference.
        Panels (b) and (c) show examples of the two TSPR query arms: in the treatment arm ($Q^B$), treated items are prioritized at the top of the ranking, while in the control arm ($Q^A$), untreated items are prioritized.}}
    \end{minipage}
    
    \label{fig:ab_vs_tspr_listings}
\end{figure}

As a baseline, we consider an item-side randomized experiment in which items are Bernoulli randomized at the listing level. Figure~\ref{fig:ab_vs_tspr_listings} contrasts this design with the Two-Sided Prioritized Ranking (TSPR) setup. In the item-side A/B test (panel~a), treated and untreated items are randomly interleaved within the same ranked list, so treated items compete directly with untreated items for user attention, generating within-list interference. TSPR instead induces structured variation in treatment exposure by simply re-ordering items within the list: In the treatment arm ($Q^B$, panel~b), treated items are promoted to the top of the ranking, whereas in the control arm ($Q^A$, panel~c), untreated items are prioritized, with placebo items buffering the two blocks. We simulate both designs and compare the resulting lift estimates.

To estimate the lift in total outcome, we extend the \cite{horvitz1952generalization} logic to the two-sided marketplace setting using item-level randomization. In this baseline configuration, randomization occurs only at the level of items. Each item $i$ is independently assigned to treatment ($Z_i = 1$) with probability $p$ or control ($Z_i = 0$) with probability $1-p$, forming the sets $T = \{i : Z_i = 1\}$ and $C = \{i : Z_i = 0\}$. Letting $y_{q,i}$ denote the outcome for item $i$ in query $q \in Q$, we estimate the mean total outcome per query under global treatment ($\mu_B$) and global control ($\mu_A$) using the estimators $\hat{\mu}_B^{IS} = \frac{1}{|Q|} \sum_{q \in Q} \sum_{i \in T} \frac{y_{q,i}}{p}$ and $\hat{\mu}_A^{IS} = \frac{1}{|Q|} \sum_{q \in Q} \sum_{i \in C} \frac{y_{q,i}}{1-p}$. Our estimand of interest is the lift $\Phi_{IS} = \frac{\mu_B}{\mu_A} - 1$. By taking the ratio of our HT estimators, the $1/|Q|$ terms cancel, yielding the item-side lift estimator:
\begin{equation}
\hat{\Phi}_{IS} = \frac{\sum_{q \in Q} \sum_{i \in T} \frac{y_{q,i}}{p}}{\sum_{q \in Q} \sum_{i \in C} \frac{y_{q,i}}{1-p}} - 1 
\label{eq:naive_ht}
\end{equation}

This baseline utilizes only item-level Bernoulli randomization and provides a simple comparison point that ignores query-level randomization.

\subsection{Performance Baseline: Cluster-Randomized Experiments}

As a second baseline, we compare our estimates to lift ratio estimates obtained from cluster-randomized experiments. Cluster randomization reduces interference bias because units within a cluster share the same treatment assignment, which limits spillover across treatment arms. However, clustering methods often exhibit substantially larger variance, as the clusters are frequently large, which effectively reduces the number of independent units of randomization. Furthermore, implementing cluster randomization requires detailed knowledge of the underlying network structure and is often costly. When it can be applied correctly, it preserves user experience coherency under our definition. For this reason, cluster-randomized experiments provide a relevant benchmark for evaluating the performance of TSPR.

To construct this baseline in our setting, we use the real search impression data from Expedia described in Section~\ref{sec:data_simulation}. Each observation consists of a property $j$ appearing in a search query $i$. Similar to the approach in \citet{holtz2024reducing}, we begin with the co-occurrence of properties across different search queries to construct our clusters. We employ Truncated SVD to find a lower-dimensional dense representation of the co-occurrence matrix and subsequently use $k$-means to cluster the resulting embeddings.

The dimension of the SVD and the number of clusters $k$ are treated as hyperparameters. We perform a grid search to identify the optimal combination based on the Modularity score. Networks with high modularity have dense connections between the nodes within modules but sparse connections between nodes in different modules. The grid search resulted in selecting 100 dimensions for SVD and 200 clusters for K-means. The resulting modularity score exceeds $0.93$, indicating that this method was able to identify highly segregated clusters of properties. 

After clustering, we randomly assign clusters to treatment with a probability $p_{\text{treat}}$. Let $T_c = \{i : \text{cluster}(i) \in \text{treated clusters}\}$ and $C_c = \{i : \text{cluster}(i) \in \text{control clusters}\}$. We compare the mean total outcome per query between treated and untreated items, and estimate lift using the same Horvitz–Thompson-style ratio structure discussed earlier for the item-side baseline:
\begin{equation}
\hat{\Phi}_{CR} = \frac{\sum_{q \in Q} \sum_{i \in T_c} \frac{y_{q,i}}{p_{\text{treat}}}}{\sum_{q \in Q} \sum_{i \in C_c} \frac{y_{q,i}}{1-p_{\text{treat}}}} - 1.
\label{eq:cluster_ht}
\end{equation}

\subsection{Main Results: Bias and Efficiency}

Before comparing estimators, we illustrate how TSPR recovers the global lift from partial-list contrasts. For the isotonic estimator, we set $\mathcal{L}_{\text{top}}$ to the largest 30\% of observed block sizes (with a minimum of two queries); results are similar for thresholds between 20\% and 50\%.

\begin{figure}[htbp]
    \centering
    \caption{Estimation of Global Lift $\Phi$ via TSPR.}
    \includegraphics[width=\textwidth]{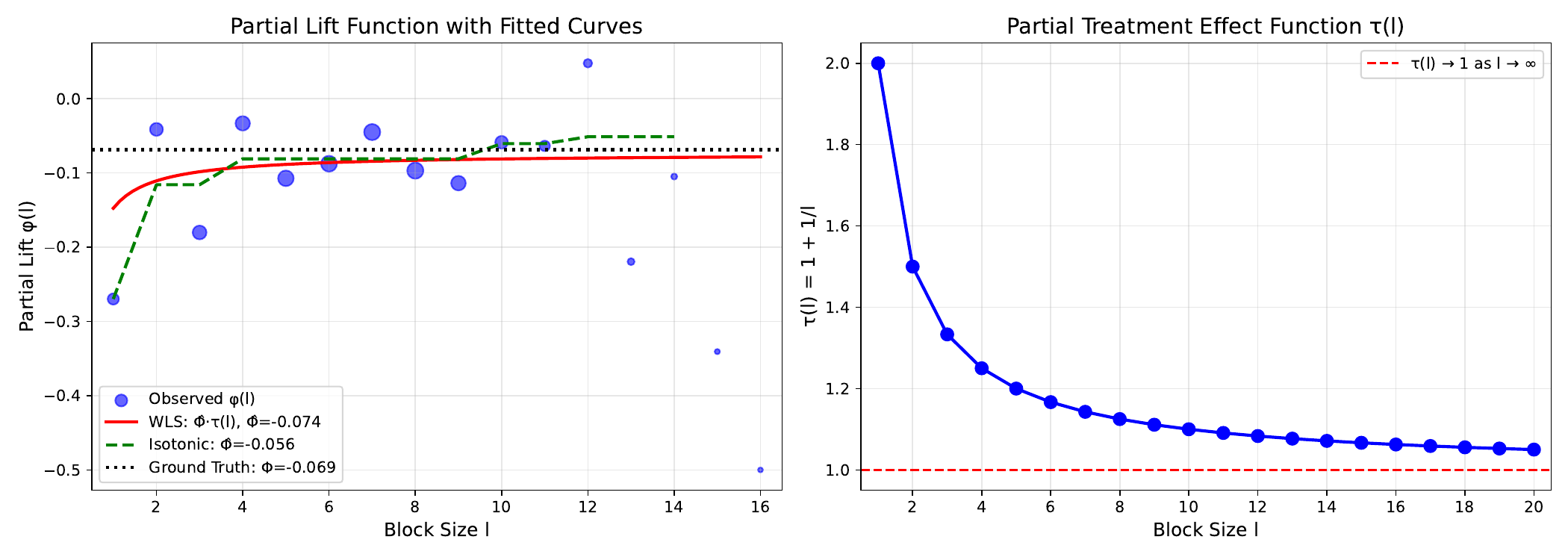}
    \parbox{0.85\textwidth}{\footnotesize{\textit{Notes:} Left: empirical partial lifts $\hat{\phi}(l)$ (blue dots; marker size proportional to precision weights $w(l)$), parametric WLS fit (red), nonparametric isotonic fit (green dashed), and ground truth $\Phi$ (dotted black). Right: the structural decay function $\tau(l) = 1 + 1/l$.}}
    \label{fig:estimation_results}
\end{figure}

Figure~\ref{fig:estimation_results} illustrates the estimation using a representative simulation. The empirical partial lifts $\hat{\phi}(l)$ are initially higher in magnitude than the global effect $\Phi$, reflecting concentrated treatment exposure at low block depths. As the block size $l$ increases, the observed lifts converge toward the ground truth. Both the parametric WLS fit and the isotonic regression capture this transition, averaging the noisy empirical points according to their query-level precision weights. The right panel highlights the structural decay of $\tau(l)$, the scaling function that maps partial lifts back to the full-treatment effect.
 
\begin{figure}[htbp]
    \centering
    \caption{Comparison of TSPR Estimators against Randomized Baselines.}
    \vspace{-2mm}
    \includegraphics[width=0.85\textwidth]{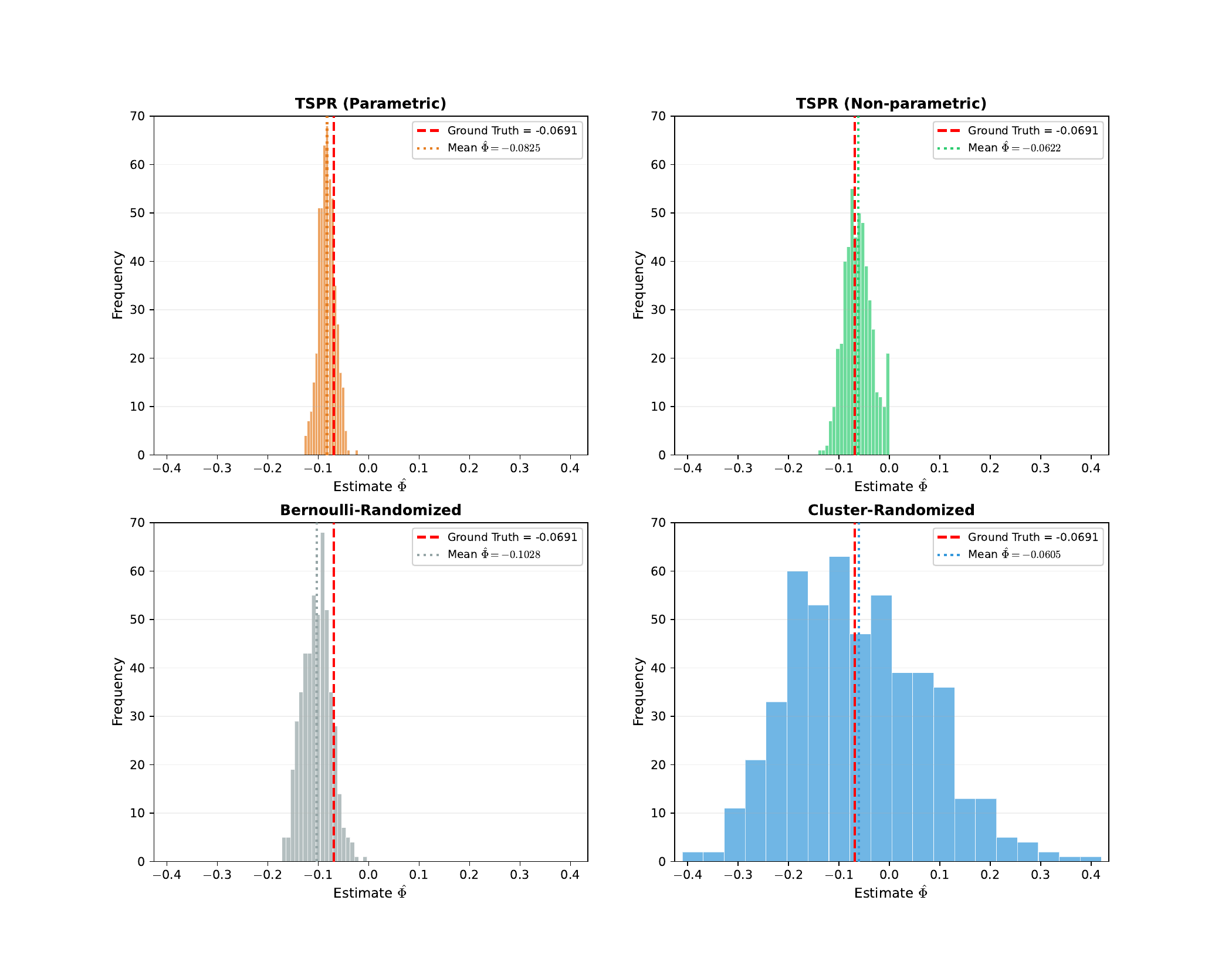}
    \parbox{\linewidth}{\footnotesize{\textit{Notes:} The figure provides histograms showing the distribution of the estimated treatment effect $\hat{\Phi}$ for parametric (WLS) and non-parametric (isotonic) TSPR, compared to Bernoulli-randomized and cluster-randomized designs.}}
    \label{fig:tspr_performance_comparison}
\end{figure}

The central results of our simulation study are summarized in Figure~\ref{fig:tspr_performance_comparison} and Table~\ref{tab:summary_stats}.

Figure~\ref{fig:tspr_performance_comparison} compares the performance of the parametric (WLS) and non-parametric (isotonic) estimators under a TSPR experiment to Bernoulli-randomized and cluster-randomized baselines. Across 500 simulation replications, TSPR delivers a substantial improvement over standard marketplace experimental designs, regardless of the estimation method. Relative to the Bernoulli-randomized (naive) baseline, TSPR effectively eliminates the severe interference bias that arises when treated and untreated items compete for attention within the same ranked list. While cluster randomization produces a mean estimate close to the ground truth (low bias), it does so at the cost of much higher variance. Specifically, the cluster-randomized design exhibits standard deviations that are 8.3$\times$ and 4.9$\times$ those of the TSPR parametric and non-parametric estimates, respectively.

Table~\ref{tab:summary_stats} quantifies these patterns in terms of bias, dispersion, and overall accuracy. The Bernoulli-randomized design exhibits substantial downward bias (mean $\hat{\Phi}=-0.1028$ vs.\ $\Phi_{true}=-0.0691$), yielding the largest bias in magnitude ($-0.0337$) and a correspondingly high RMSE (0.0438). Cluster randomization largely removes bias (bias $=0.0086$), but its performance is dominated by extreme variability (empirical SD $=0.1345$), producing by far the worst RMSE (0.1348). In contrast, both TSPR estimators achieve a markedly better bias–variance tradeoff. The parametric TSPR estimator attains the lowest RMSE (0.0210) and the smallest empirical SD (0.0162), despite a modest bias ($-0.0133$). The non-parametric TSPR estimator exhibits a very low bias (bias $=0.0069$) while maintaining low variance (SD $=0.0275$) and a low RMSE (0.0283). Overall, the table shows that TSPR improves accuracy relative to Bernoulli randomization by sharply reducing interference-driven bias, while avoiding the prohibitive variance costs of clustering.

\begin{table}[h]
\centering
\caption{Summary Statistics of Estimator Performance ($\Phi_{true} = -0.0691$)}
\label{tab:summary_stats}
\begin{tabular}{lccccc}
\hline
\textbf{Estimator} & \textbf{Mean $\hat{\Phi}$} & \textbf{Bias} & \textbf{RMSE} & \textbf{Empirical SD} \\ \hline
Bernoulli-Randomized        & -0.1028 & -0.0337 & 0.0438 & 0.0279 \\
Cluster-Randomized    & -0.0605 & 0.0086  & 0.1348 & 0.1345 \\
TSPR (Parametric)     & -0.0825 & -0.0133 & 0.0210 & 0.0162 \\
TSPR (Non-parametric) & -0.0622 & 0.0069  & 0.0283 & 0.0275 \\ \hline
\end{tabular}
\end{table}

We assess sensitivity to the treatment probability $p$. While our main results use $p=0.25$, we stress-test TSPR over $p \in [0.10, 0.45]$, which corresponds to placebo buffer sizes ranging from 80\% to 10\%. As shown in Appendix~\ref{app:sensitivity_p}, TSPR remains robust across this range, consistently delivering substantially lower bias and variance than the Bernoulli-randomized baseline.

Bootstrap confidence intervals for the TSPR non-parametric estimator achieve 93\% empirical coverage at the nominal 95\% level, indicating well-calibrated uncertainty quantification (Appendix~\ref{app:coverage}).

\section{Discussion}
\label{sec:discussion}

The Bernoulli-randomized item-side estimator is unbiased only when each item's outcome is independent of all other items in the list. In marketplace search, this no-interference condition fails: users substitute across items, so that booking one option crowds out others within the same query. Position bias amplifies the resulting distortion, because the items most affected by substitution are those competing in high-attention ranks where the bulk of user engagement occurs. When interference and position bias act together, the naive estimator systematically overstates the magnitude of the treatment effect, conflating the causal impact of treatment with the reallocation of demand across treated and untreated items within the same list.

TSPR addresses this problem by leveraging position bias as the source of identifying variation. By reordering ranked lists so that one query group receives concentrated exposure to treated items and the other to untreated items, TSPR creates meaningful differences in aggregate outcomes across arms without requiring interference to be absent. The design reduces bias from within-list substitution, and remains effective precisely in the environments where naive item-level experiments fail most severely. If user behavior were order-invariant, with clicking and booking depending only on the set of utilities $\{v_{qi}\}$ and not their positions, then permuting the ranking would not change outcomes, and TSPR would have no identifying variation. TSPR is therefore most valuable in environments where position bias is strong, a condition that is well documented in search and recommendation settings.

Cluster randomization provides an alternative by assigning groups of related items to the same treatment status, eliminating within-cluster spillovers. When cluster boundaries align with the true interference structure, this approach can achieve low bias, as in our simulation where SVD-based embedding followed by
k-means yielded modularity above 0.93. The efficiency cost, however, is substantial: with roughly 200 clusters instead of 20{,}000 query-level randomization units in TSPR, the effective sample size falls by about two orders of magnitude, producing roughly 8 times the standard deviation of TSPR’s parametric estimator. Moreover, clean clusters are difficult to construct in practice. Substitution patterns are often diffuse and context-dependent, and misspecified boundaries reintroduce the very spillovers clustering is meant to remove. TSPR avoids this fragility and does not depend on knowing the interference network.

\section{Conclusion}
\label{sec:conclusion}

This paper introduces Two-Sided Prioritized Ranking (TSPR), an
experimental design for item-side interventions in online marketplaces that maintains price parity and full catalog access while addressing interference through position-based exposure variation.  In simulations calibrated to hotel search data, TSPR substantially reduces bias relative to Bernoulli-randomized A/B tests and achieves an order-of-magnitude reduction in standard deviation compared to cluster randomization, even when clusters are cleanly defined.

TSPR can be implemented by adjusting ranking priorities within an existing recommender system, making it straightforward to deploy on platforms that already support re-ranking logic. The design avoids the operational and data burdens that often accompany marketplace experimentation, such as user-level pricing changes, catalog partitioning, or explicitly modeling the interference network. Although TSPR does not target the global treatment effect without bias, it offers a strongly favorable bias–variance tradeoff in practice: relative to Bernoulli randomization it substantially reduces both bias and variance, and relative to cluster randomization it delivers far higher precision.

TSPR is best suited to settings with strong position bias, slack supply, and short experiment horizons that limit dynamic feedback. The identifying assumptions of treatment-attention separability and symmetric re-ranking distortion are plausible in many marketplace settings, but they can fail if the treatment meaningfully changes user engagement or if high treatment probabilities induce ranking perturbations that users do not tolerate. Practitioners should assess these risks using pre-experiment diagnostics and the sensitivity analyses reported in the appendix.

Natural extensions include adapting TSPR to supply-constrained settings with binding capacity constraints, extending it to continuous or multi-arm treatments, and allowing for user-side interference across queries. More broadly, the results underscore that platform structure, here ranked attention, can provide identifying variation even under interference. We view the coherency constraints that motivate TSPR as realistic operational requirements, and a useful guide for methodological development in platform experimentation.

\clearpage

\bibliographystyle{agsm}   
\bibliography{ref}

\clearpage

\appendix
\section{Appendix}

\subsection{Diagnostics for Identifying Assumptions}
\label{app:empirical_support_assumptions}

We provide simulation-based diagnostics for the two main identifying assumptions of the TSPR framework: Assumption~\ref{ass:separability} (Attention and Treatment Separability) and Assumption~\ref{ass:symmetric_distortion} (Symmetric Distortion). We refer to experiments in which the TSPR re-ranking mechanism is applied but no treatment effect is imposed ($\tau=0$) as \emph{dummy experiments}. In such experiments, any systematic difference in outcomes across arms must arise from the design rather than from treatment, providing a direct test of the assumptions.

\begin{figure}[htbp]
    \centering
    \includegraphics[width=0.65\linewidth]{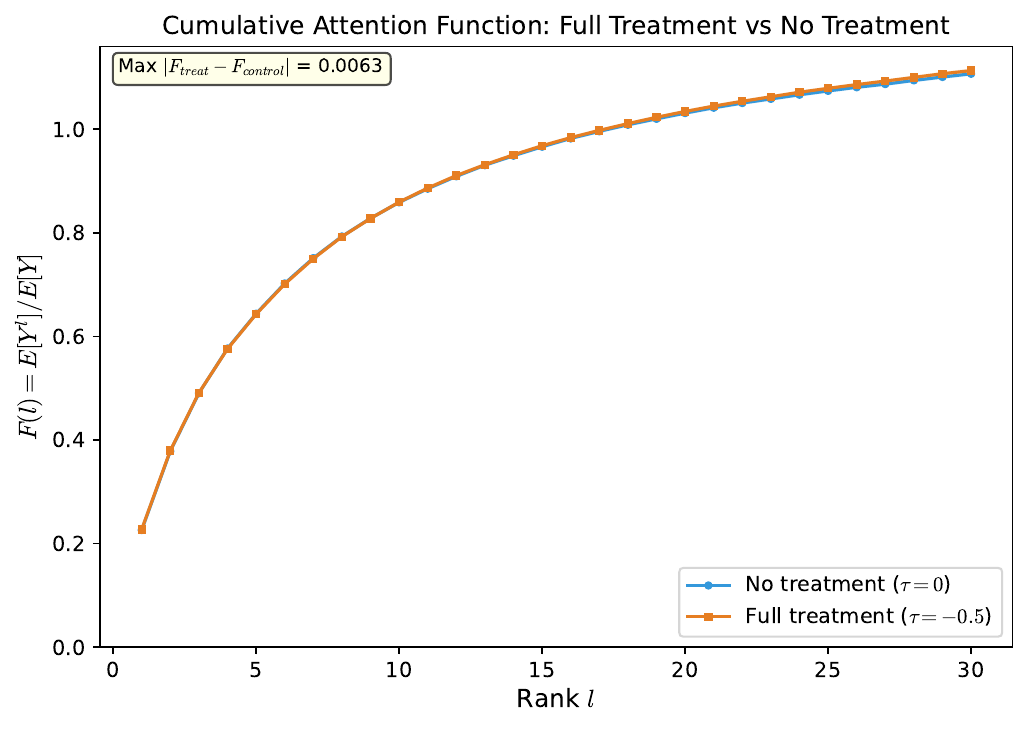}
    \caption{
    Cumulative attention functions under full-treatment ($\tau=-0.5$) and no-treatment ($\tau=0$) counterfactuals, each averaged over 30 simulation draws under the original ranking.
    Both curves are increasing and concave, consistent with Definition~\ref{def:attention}.
    The near-complete overlap ($\max |F_{\text{treat}} - F_{\text{control}}| = 0.006$) indicates that treatment scales outcomes without altering the shape of attention, supporting Assumption~\ref{ass:separability}.
    }
    \label{fig:attention_function}
\end{figure}

\paragraph{Separability (Assumption~\ref{ass:separability}).}
Assumption~\ref{ass:separability} requires that treatment scales the level of outcomes without altering the shape of the attention function $F(l)$.
To assess this, we simulate two counterfactual worlds under the platform's original (unmodified) ranking: one with no treatment and one with full treatment, each averaged over 30 independent draws.
Figure~\ref{fig:attention_function} plots the normalized cumulative attention function $F(l)=\mathbb{E}[Y^{l}]/\mathbb{E}[Y]$ as a function of rank $l$ in both scenarios.
Both curves are increasing and concave, consistent with Definition~\ref{def:attention}.
The two profiles are nearly indistinguishable: the maximum pointwise absolute difference is $\max_l |F_{\text{treat}}(l) - F_{\text{control}}(l)| = 0.006$, confirming that treatment shifts the level of outcomes without meaningfully altering how attention is allocated across ranks.

\paragraph{Symmetric distortion (Assumption~\ref{ass:symmetric_distortion}).}
Assumption~\ref{ass:symmetric_distortion} requires that the TSPR re-ranking induces the same expected attention attenuation in both experimental arms: $d_A(l;p) = d_B(l;p)$ for all ranks~$l$.
We present three complementary diagnostics.

\begin{figure}[htbp]
    \centering
    \includegraphics[width=0.75\linewidth]{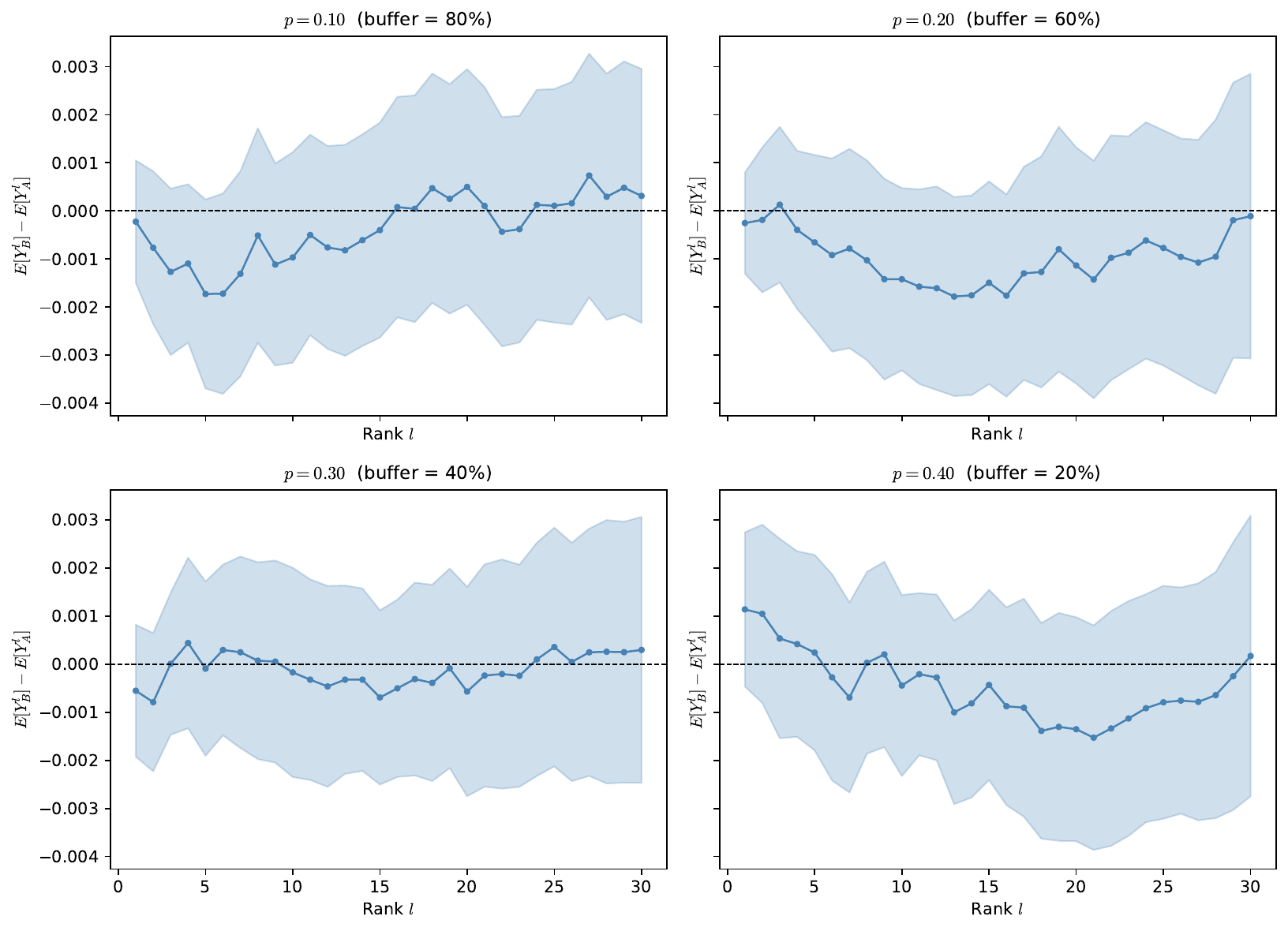}
    \caption{Dummy experiment diagnostic for symmetric distortion.
    The figure plots $\mathbb{E}[Y^{l}]_B - \mathbb{E}[Y^{l}]_A$ by rank $l$ in dummy experiments where treatment labels are assigned but no treatment is applied ($\tau=0$), averaged over 50 draws.
    Panels correspond to different treatment assignment probabilities $p \in \{0.10, 0.20, 0.30, 0.40\}$.
    Shaded regions denote pointwise 95\% confidence intervals.
    }
    \label{fig:AA_symmetric_distortion_stress_test}
\end{figure}

\medskip
\noindent\textit{(i) Dummy experiment: partial-outcome differences under null treatment.}
We conduct dummy experiments in which the TSPR re-ranking is applied but no treatment effect is imposed ($\tau=0$).
Under symmetric distortion, $\mathbb{E}[Y^{l}_B] - \mathbb{E}[Y^{l}_A] = 0$ for all~$l$.
Figure~\ref{fig:AA_symmetric_distortion_stress_test} plots these differences for treatment assignment probabilities $p \in \{0.10, 0.20, 0.30, 0.40\}$, each averaged over 50 independent draws.
Across all values of $p$ and all ranks, the differences remain small (maximum absolute mean difference $< 0.002$, relative to baseline partial outcomes of order 0.7), fluctuate around zero, and exhibit no systematic pattern.
The widening confidence intervals at deeper ranks reflect the cumulative nature of the outcome rather than asymmetric attenuation.

\begin{figure}[htbp]
    \centering
    \includegraphics[width=0.8\textwidth]{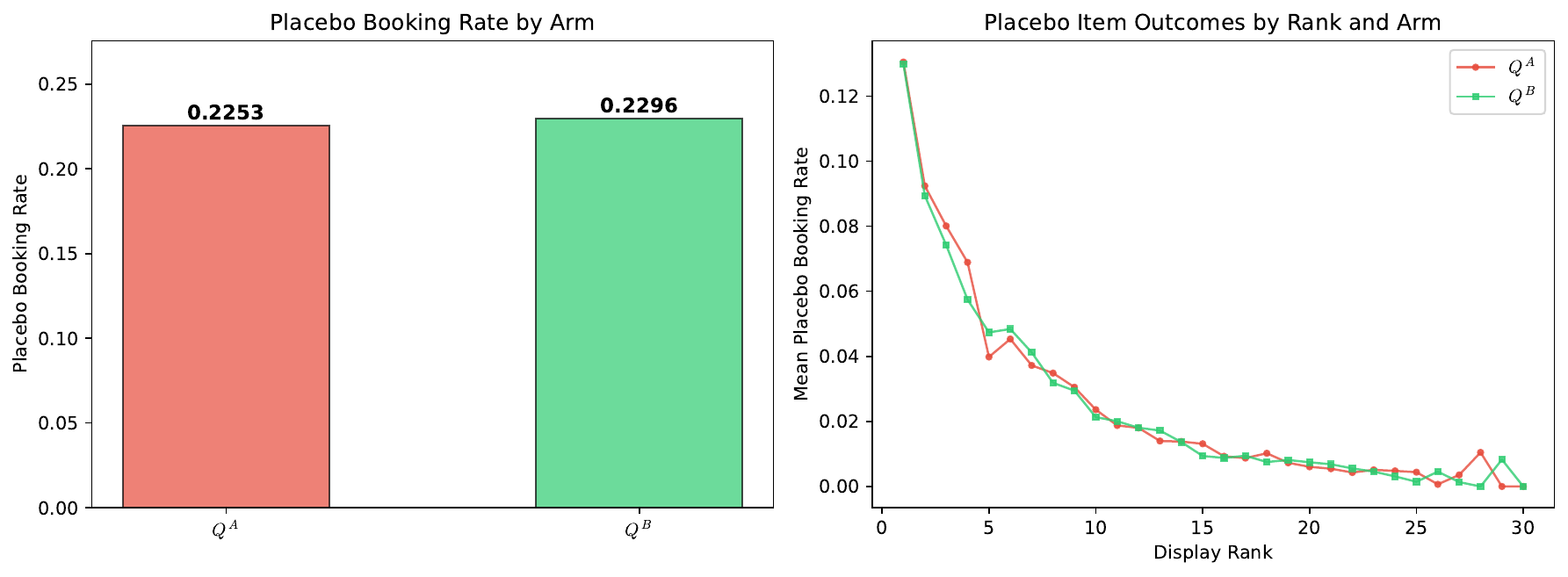}
    \caption{
    Placebo outcome balance across experimental arms under zero treatment with $p=0.25$.
    Left: placebo booking rates by arm.
    Right: mean placebo booking rate by display rank and arm.}
    \label{fig:placebo_balance}
\end{figure}

\medskip
\noindent\textit{(ii) Placebo outcome balance.}
Placebo items receive no treatment in either arm and occupy the middle block (priority~2) under TSPR.
Any difference in placebo outcomes between arms would therefore indicate asymmetric distortion in the re-ranking mechanism itself.
In a dummy experiment with $p=0.25$, we compare the fraction of queries in which at least one placebo item is booked across arms.
The placebo booking rates are 22.5\% ($Q^A$, $n=9{,}973$) and 23.0\% ($Q^B$, $n=9{,}996$), with a two-proportion $z$-test yielding $z=0.84$ ($p=0.40$).
Figure~\ref{fig:placebo_balance} confirms this balance, showing that rank-specific placebo booking rates are nearly identical across arms throughout the listing.

\begin{figure}[htbp]
    \centering
    \includegraphics[width=0.65\linewidth]{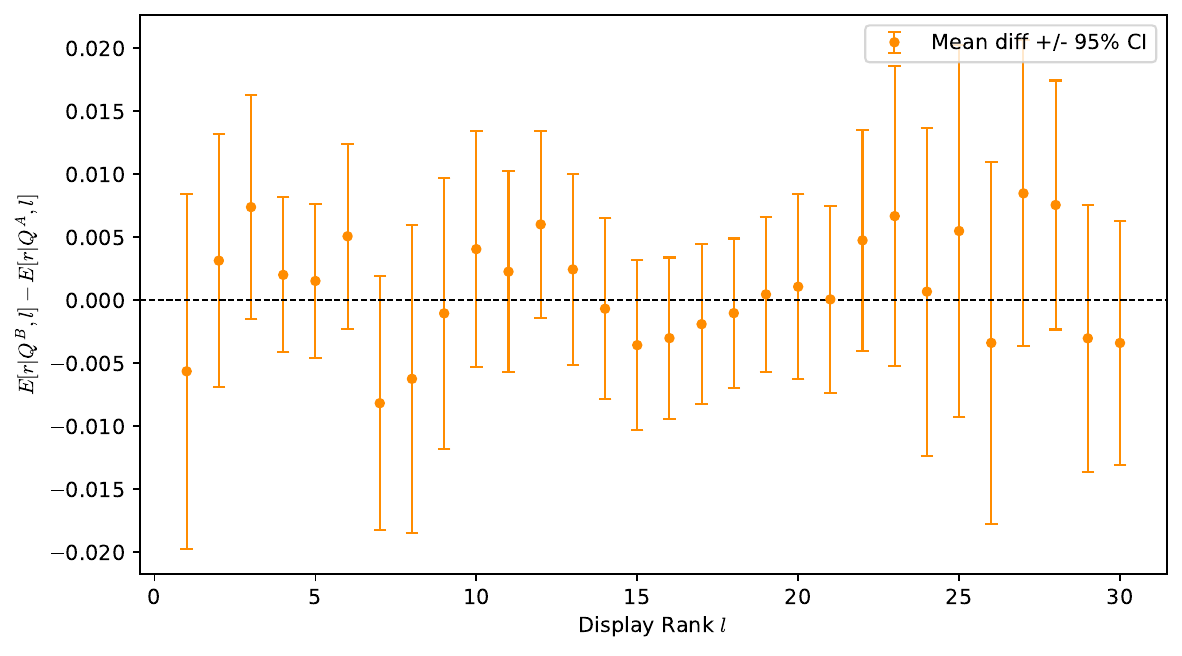}
    \caption{Relevance Balance Check. Rank-specific difference in mean relevance between arms $Q^B$ and $Q^A$, in 50 simulations with $p=0.25$. The zero line corresponds to symmetric distortion.}
    \label{fig:symmetric_distortion_diagnostic}
\end{figure}

\medskip
\noindent\textit{(iii) Relevance balance by rank.}
Because items are randomly assigned to treatment groups independently of their baseline relevance, the expected relevance at each display rank should be equal across arms.
Figure~\ref{fig:symmetric_distortion_diagnostic} plots the rank-specific difference in mean pre-treatment relevance, $\mathbb{E}[r \mid Q^B, l] - \mathbb{E}[r \mid Q^A, l]$, averaged over 50 independent simulation draws.
The differences are tightly centered around zero with no systematic rank-dependent pattern, confirming that the block-swap mechanism does not create quality imbalances between arms.

\begin{figure}[htbp]
    \centering
    \includegraphics[width=0.65\textwidth]{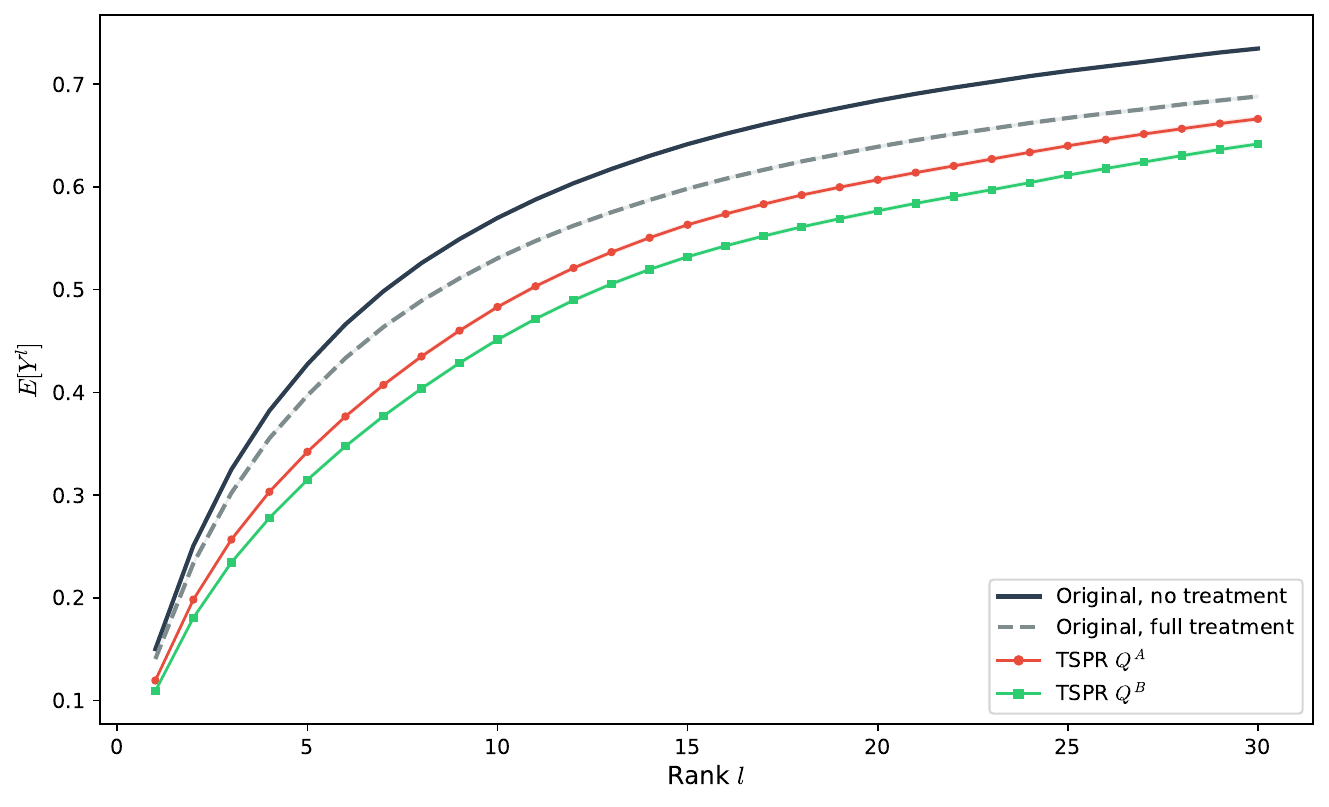}
    \caption{
    Mean partial outcomes $\mathbb{E}[Y^l]$ by rank under four scenarios, each averaged over 50 simulation draws.
    The two solid dark lines show the original ranking under no treatment and full treatment.
    The red and green lines show TSPR arms $Q^A$ and $Q^B$ with treatment applied at $p=0.25$.
    The TSPR curves lie below the original-ranking baselines, reflecting re-ranking attenuation. The gap between $Q^A$ and $Q^B$ reflects differential treatment exposure, the identifying variation exploited by TSPR.
    }
    \label{fig:par_y}
\end{figure}

\medskip
\noindent\textit{Joint diagnostic: partial outcomes under TSPR versus counterfactuals.}
As a joint check of both Assumptions~\ref{ass:separability} and \ref{ass:symmetric_distortion}, Figure~\ref{fig:par_y} plots mean partial outcomes $\mathbb{E}[Y^l]$ under four scenarios: original ranking with no treatment, original ranking with full treatment, and the two TSPR arms ($Q^A$ and $Q^B$) with treatment applied at $p=0.25$.
The TSPR arms lie below the original-ranking curves, confirming that re-ranking attenuates outcomes as formalized by the distortion function $d(l;p)$ in Assumption~\ref{ass:multiplicative_distortion}.

The gap between the $Q^A$ and $Q^B$ curves reflects the differential treatment exposure that TSPR exploits for identification, not asymmetric distortion: $Q^B$ promotes treated items (which reduce bookings under $\tau < 0$) while $Q^A$ promotes untreated items.
Both TSPR curves track the same concave shape as the counterfactuals, consistent with multiplicative rather than shape-altering distortion.

\clearpage
\subsection{Booking Model Comparison}
\label{appendix:booking_model_comparison}

We compare two specifications of the booking model: a standard multinomial logit (MNL) that conditions only on item utility, and the cascade model defined in Equation~\eqref{eq:booking_model} that adds a click-order effect.  Both models are estimated by maximum likelihood on the same sample of clicked items.

The MNL model is
\begin{equation*}
P(\text{book}_{ik} \mid C_i) \;=\;
  \frac{\exp\!\big(\gamma_1 \, v_{ik} + \gamma_0\big)}
       {1 + \sum_{k' \in C_i} \exp\!\big(\gamma_1 \, v_{ik'} + \gamma_0\big)},
\end{equation*}
which is nested within the cascade specification by the restriction $\gamma_2 = 0$. Table \ref{tab:booking_params_mnl} provides the parameters for a standard Multinomial Logit (MNL) specification without click-order effects. This model was used as a baseline for the likelihood ratio test mentioned in Section \ref{subsection:booking_model}.

\begin{table}[ht]
\centering
\caption{Standard MNL Booking Parameters (Baseline)}
\begin{tabular}{llr}
\hline
Parameter & Variable & Estimate \\
\hline
$\gamma_1$ & Latent Utility ($v$) & 0.4472 \\
$\gamma_0$ & Intercept & 0.2376 \\
\hline
\end{tabular}
\label{tab:booking_params_mnl}
\end{table}

Table~\ref{tab:booking_comparison} reports the log-likelihoods and the likelihood ratio test.

\begin{table}[h]
  \centering
  \caption{Booking model comparison: MNL vs.\ Cascade.}
  \label{tab:booking_comparison}
  \begin{tabular}{lcc}
    \toprule
    Model & Log-likelihood & Parameters \\
    \midrule
    Standard MNL          & $-49{,}251.4$ & 2 \\
    Cascade (click order) & $-49{,}205.3$ & 3 \\
    \midrule
    \multicolumn{3}{l}{Likelihood ratio statistic: $\chi^2 = 92.1$ \quad ($p < 0.001$, $df = 1$)} \\
    \bottomrule
  \end{tabular}
\end{table}

The cascade model achieves a significantly higher log-likelihood with the addition of a single parameter (click order), yielding a likelihood ratio statistic of $92.1$ on one degree of freedom ($p < 0.001$).  The estimated click-order coefficient is negative ($\hat{\gamma}_2 < 0$), confirming that items clicked earlier are more likely to be booked.  We therefore adopt the cascade specification for all main results.

\newpage
\subsection{Coverage Analysis}
\label{app:coverage}

This appendix provides an assessment of confidence-interval
calibration for the two estimators used to estimate the total lift in the TSPR experimental design.  For each Monte Carlo
run, standard errors are computed via the nonparametric query-level
bootstrap with $B = 50$ resamples drawn with replacement.  Normal-approximation
95\% confidence intervals are constructed as
$\hat\Phi \pm 1.96\,\widehat{\mathrm{SE}}$.

\paragraph{Decomposing under-coverage.}
Confidence-interval coverage can fail for two distinct reasons:
(i)~the bootstrap standard error underestimates the true sampling
variability of the estimator (\emph{SE miscalibration}), or (ii)~the
estimator is biased so that intervals are systematically shifted away
from the truth (\emph{bias-induced miss}).  We quantify these two
channels with the following diagnostics:

\begin{itemize}[nosep]
  \item \textbf{SE-to-SD ratio} $= \overline{\widehat{\mathrm{SE}}} \;/\; \mathrm{SD}(\hat\Phi)$:
        the mean bootstrap SE divided by the empirical standard
        deviation of the point estimates across Monte Carlo runs.
        A ratio of 1.0 indicates perfect SE calibration.
  \item \textbf{Bias-to-SD ratio} $= |\mathrm{Bias}| \;/\; \mathrm{SD}(\hat\Phi)$:
        the absolute bias normalized by the empirical SD.  Values
        above $\approx 0.5$ indicate that bias is a meaningful
        contributor to under-coverage.
\end{itemize}

\begin{table}[h]
\centering
\caption{Coverage Diagnostics ($\Phi_{\mathrm{true}} = -0.0691$, 500 Monte Carlo runs)}
\label{tab:coverage_diagnostics}
\begin{tabular}{lccccc}
\toprule
\textbf{Estimator}
  & \textbf{SE/SD}
  & \textbf{$|\text{Bias}|$/SD}
  & \textbf{95\% Coverage}
  & \textbf{Primary Driver} \\
\midrule
TSPR (Parametric)
  & 0.89 & 0.82 & 80.0\% & Primarily bias \\
TSPR (Non-parametric)
  & 0.99 & 0.25 & 92.6\% & Well calibrated \\
\bottomrule
\end{tabular}
\end{table}

The parametric TSPR estimator achieves good SE calibration (SE/SD~$= 0.89$).
The residual under-coverage (80.0\% vs.\ 95\%) is driven almost
entirely by the estimator's bias: the parametric form
$\tau(l) = 1 + \frac{1}{l}$ does not perfectly match the true shape of the
partial treatment-effect function, inducing a systematic shift of
$\approx 0.82$ empirical standard deviations.
This estimator nevertheless achieves the lowest RMSE of all four
designs, making it the preferred choice when point-estimation
accuracy is the primary objective.

The isotonic regression estimator achieves near-perfect SE
calibration (SE/SD~$= 0.99$) and low bias (ratio~0.25), yielding
92.6\% coverage, close to the nominal 95\% level.  This makes
it the preferred choice when valid inferential coverage is required,
for instance when the goal is to detect whether a treatment effect
is statistically significant.

The two TSPR estimators offer complementary strengths: the parametric variant minimizes mean-squared error and is best suited for point estimation, while the non-parametric variant provides well-calibrated confidence intervals for hypothesis testing. Both represent a substantial improvement over conventional designs, which suffer from either severe bias (Bernoulli) or inflated variance (cluster randomization).

\newpage
\subsection{Sensitivity to Treatment Probability}
\label{app:sensitivity_p}

We examine the robustness of our estimator to the choice of treatment probability $p$, which determines the allocation of items across three experimental groups: treated ($p$), untreated ($p$), and the placebo buffer ($1-2p$). In our main specification, we set $p = 0.25$, yielding equal 25\% shares for treated and untreated items with a 50\% placebo buffer. We vary $p$ from 0.10 to 0.45, corresponding to buffer sizes ranging from 80\% down to 10\%.

\begin{figure}[htbp]
    \centering
    \includegraphics[width=\textwidth]{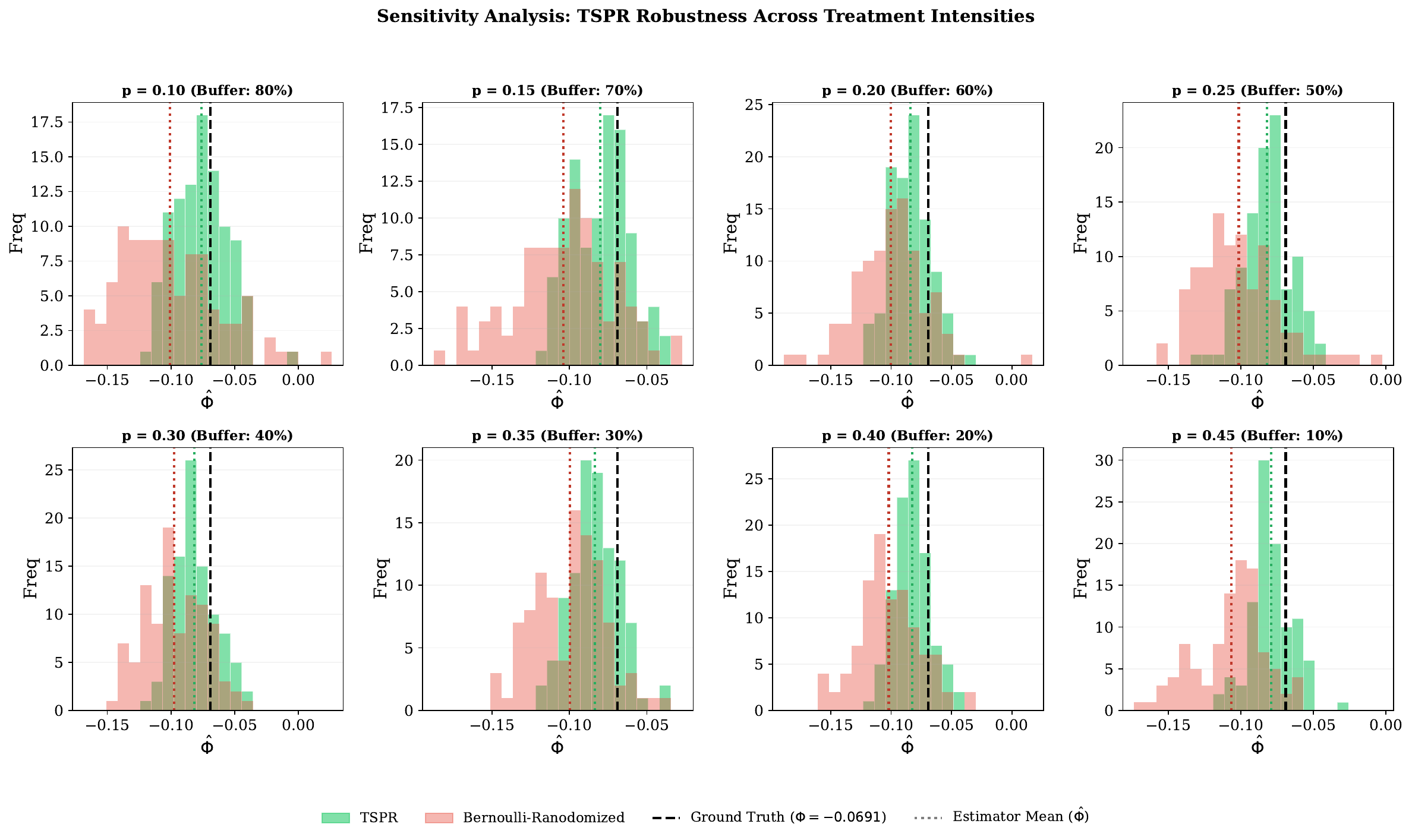}
    \caption{Distribution of Estimates ($\hat{\Phi}$) across different treatment probabilities ($p$). The dashed black line represents the ground truth value of -0.0691. TSPR (green) consistently exhibits lower spread and better centering on the ground truth compared to the Naive estimator (red) across all buffer sizes.}
    \label{fig:sensitivity_p_hists}
\end{figure}

\begin{figure}[htbp]
    \centering
    \begin{subfigure}[b]{0.48\textwidth}
        \centering
        \includegraphics[width=\textwidth]{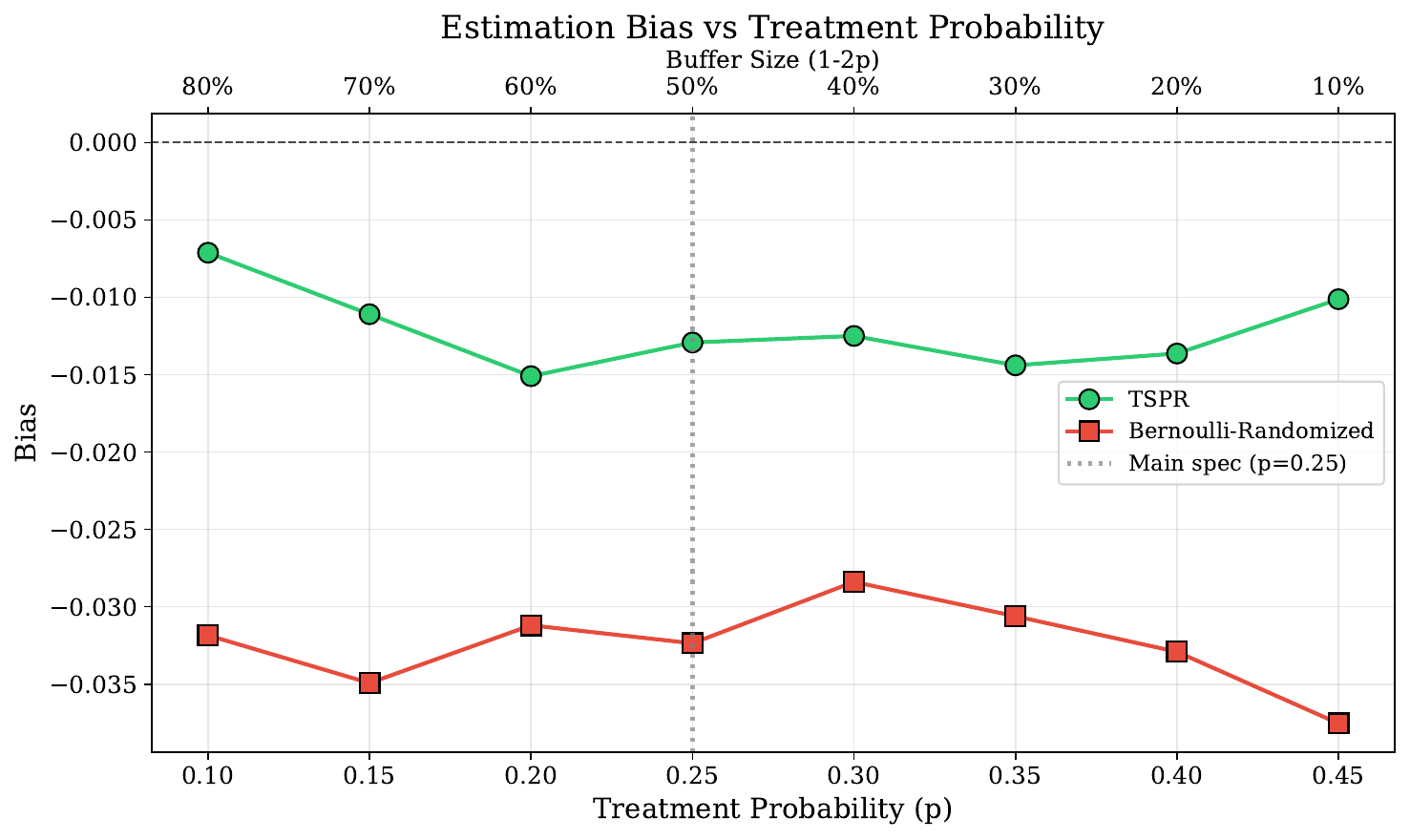}
        \caption{Estimation Bias}
        \label{fig:sensitivity_p_bias}
    \end{subfigure}
    \hfill
    \begin{subfigure}[b]{0.48\textwidth}
        \centering
        \includegraphics[width=\textwidth]{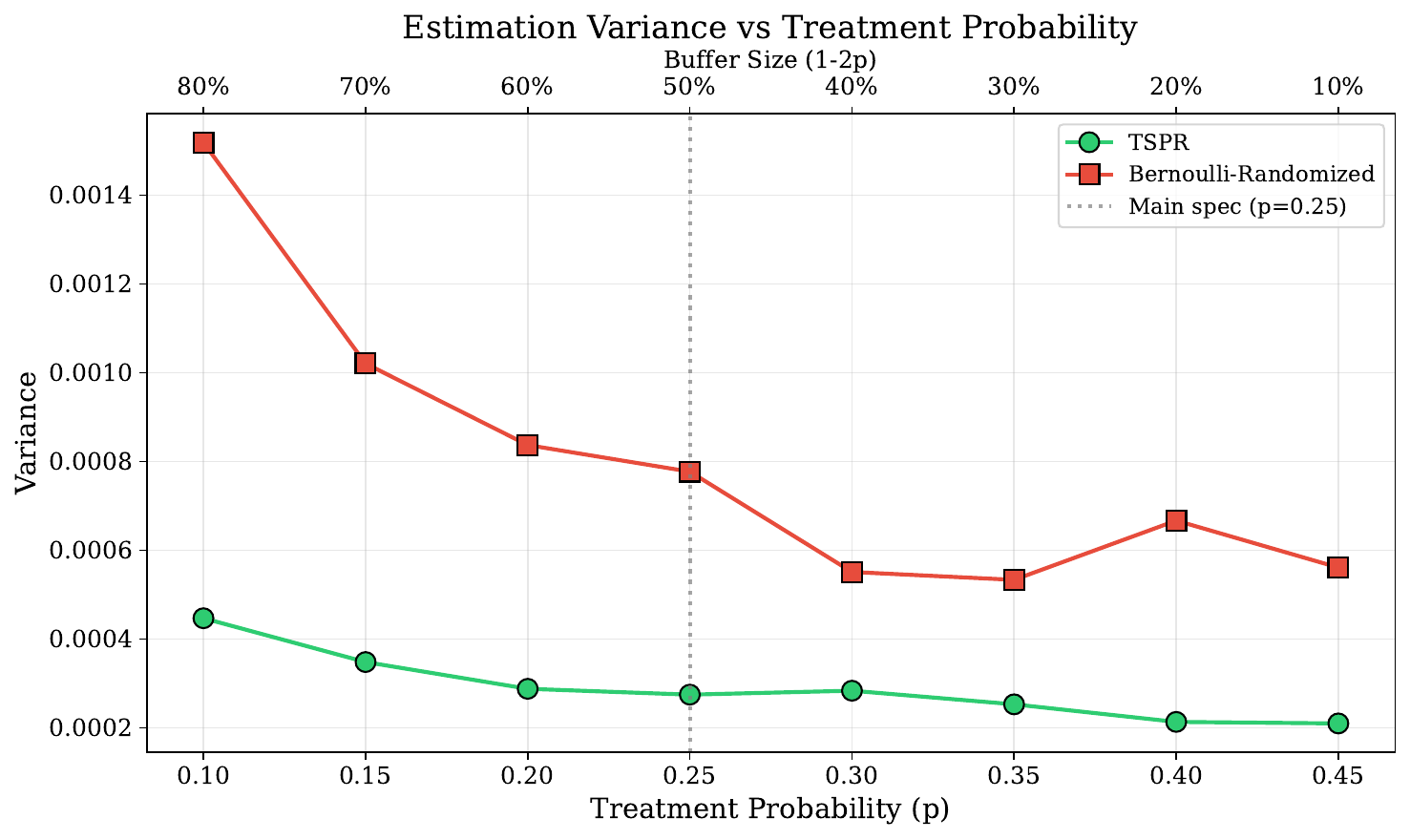}
        \caption{Estimation Variance}
        \label{fig:sensitivity_p_variance}
    \end{subfigure}
    \caption{Sensitivity to treatment probability $p$. Panel (a) shows estimation bias and Panel (b) shows estimation variance for TSPR and the Bernoulli-Randomized estimator across different values of $p$. The buffer size (1-2$p$) is shown on the secondary axis on top. The vertical dotted line indicates our main specification ($p = 0.25$). TSPR consistently achieves lower bias and variance than the naive estimator across all tested configurations.}
    \label{fig:sensitivity_p}
\end{figure}

Figure~\ref{fig:sensitivity_p_hists} compares the distribution of lift estimates ($\hat{\Phi}$) for TSPR versus the naive item-side A/B test across eight different levels of $p$. In every scenario, the TSPR distribution is more tightly centered around the ground truth ($\Phi = -0.0691$). As shown in Figure~\ref{fig:sensitivity_p}, TSPR consistently outperforms the naive experiment in both bias and variance.

TSPR achieves significantly lower bias than the naive estimator across all tested values of $p$. The naive estimator exhibits substantial negative bias, with an absolute relative bias ranging from 41.9\% to 55.3\% of the ground truth. This magnitude reflects the severe interference that occurs when treated and untreated items compete for finite user attention within the same ranking list. In contrast, TSPR dramatically reduces this error, maintaining an absolute relative bias between 11.0\% and 22.6\%. By decoupling the competition through prioritized ranking, TSPR yields estimates that are consistently closer to the true treatment effect, regardless of the specific allocation.

The variance results similarly favor TSPR, which exhibits lower estimation variance than the naive approach across all values of $p$. Both estimators show a general downward trend in variance as $p$ increases. For Bernoulli, the variance decline comes from better balance between treated and control groups, while for TSPR it comes from larger treated/untreated pools.

These results confirm that TSPR is robust to the choice of $p$ and provides reliable, interference-corrected estimates across a wide range of experimental configurations.

\end{document}